\documentclass[12pt]{article}

\usepackage{epstopdf,amsfonts,amsmath}
\usepackage{graphicx}
\usepackage{tikz}
\usepackage{stmaryrd}
\usepackage{etoolbox}
\usepackage{mathdots} 
\DeclareGraphicsExtensions{.eps}

\newcommand\encadremath[1]{\vbox{\hrule\hbox{\vrule\kern8pt
\vbox{\kern8pt \hbox{$\displaystyle #1$}\kern8pt}
\kern8pt\vrule}\hrule}}
\def\enca#1{\vbox{\hrule\hbox{
\vrule\kern8pt\vbox{\kern8pt \hbox{$\displaystyle #1$}
\kern8pt} \kern8pt\vrule}\hrule}}

\newcommand\figureframex[3]{
\begin{figure}[bth]
\hrule\hbox{\vrule\kern8pt
\vbox{\kern8pt \vbox{
\begin{center}
{\mbox{\epsfxsize=#1.truecm\epsfbox{#2}}}
\end{center}
\caption{#3}
}\kern8pt}
\kern8pt\vrule}\hrule
\end{figure}
}
\newcommand\figureframey[3]{
\begin{figure}[bth]
\hrule\hbox{\vrule\kern8pt
\vbox{\kern8pt \vbox{
\begin{center}
{\mbox{\epsfysize=#1.truecm\epsfbox{#2}}}
\end{center}
\caption{#3}
}\kern8pt}
\kern8pt\vrule}\hrule
\end{figure}
}

\makeatletter
\@addtoreset{equation}{section}
\makeatother
\newtheorem{theorem}{Theorem}[section]
\newtheorem{conjecture}{Conjecture}[section]
\newtheorem{remark}{Remark}[section]
\newtheorem{proposition}{Proposition}[section]
\newtheorem{lemma}{Lemma}[section]
\newtheorem{corollary}{Corollary}[section]
\newtheorem{definition}{Definition}[section]
\newtheorem{assumption}{Assumption}
\def\br{\begin{remark}\rm\small}
\def\er{\end{remark}}
\def\bt{\begin{theorem}}
\def\et{\end{theorem}}
\def\bd{\begin{definition}}
\def\ed{\end{definition}}
\def\bp{\begin{proposition}}
\def\ep{\end{proposition}}
\def\bl{\begin{lemma}}
\def\el{\end{lemma}}
\def\bc{\begin{corollary}}
\def\ec{\end{corollary}}
\def\beaq{\begin{eqnarray}}
\def\eeaq{\end{eqnarray}}
\newcommand{\proof}[1]{{\noindent \bf Proof  :}\par
{#1} $\square$}

\newcommand{\td}{\tilde}

\newcommand{\beq}{\begin{equation}}
\newcommand{\eeq}{\end{equation}}
\newcommand{\beqq}{\begin{equation*}}
\newcommand{\eeqq}{\end{equation*}}
\newcommand{\bea}{\begin{eqnarray}}
\newcommand{\eea}{\end{eqnarray}}
\newcommand{\beaa}{\begin{eqnarray*}}
\newcommand{\eeaa}{\end{eqnarray*}}

\newcommand{\Tr}{\operatorname{Tr}}

\newcommand{\Lieg}{{\mathfrak g}}
\newcommand{\Lieh}{{\mathfrak h}}

\newcommand{\Tau}{{\mathfrak T}}
\newcommand{\W}{{\cal W}}
\newcommand{\spcurve}{{\cal S}}
\newcommand{\curve}{{\Sigma}}

\newcommand{\x}{{\mathrm x}}
\newcommand{\y}{{\mathrm y}}
\newcommand{\s}{{\mathrm s}}

\renewcommand{\W}{{\mathcal W}}

\newcommand{\Res}{\mathop{\,\rm Res\,}}
\usepackage[pdftex]{hyperref}
\hypersetup{colorlinks,urlcolor=magenta,citecolor=red,linkcolor=blue,filecolor=black}

\begin{document}

\hfill
\sloppy

\hfill IPHT-t16/091\\
\indent \hfill CRM-3355

\addtolength{\baselineskip}{0.20\baselineskip}
\begin{center}
\vspace{1cm}

{\Large \bf {Integrable differential systems of topological type and reconstruction by the topological recursion}}

\vspace{1cm}

{Rapha\"el Belliard}$^1$,
{Bertrand Eynard}$^{1,2}$,
{Olivier Marchal}$^3$

\vspace{5mm}
$^1$\ Institut de physique th\'eorique, Universit\'e Paris Saclay, 
\\
CEA, CNRS, F-91191 Gif-sur-Yvette, France
\vspace{5mm}
\\
$^2$\ Centre de recherches math\'ematiques, Universit\'e de Montr\'eal, Canada
\vspace{5mm}
\\
$^3$\  Universit\'{e} de Lyon, CNRS UMR 5208, Universit\'{e} Jean Monnet,
\\
Institut Camille Jordan, France
\vspace{5mm}
\\
\end{center}

\vspace{1cm}
\begin{center}
{\bf Abstract :}
Starting from a $d\times d$ rational Lax pair system of the form $\hbar \partial_x \Psi= L\Psi$ and $\hbar \partial_t \Psi=R\Psi$ we prove that, under certain assumptions (genus $0$ spectral curve and additional conditions on $R$ and $L$),  the system satisfies the ``topological type property''. 
A consequence is that the formal $\hbar$-WKB expansion of its determinantal correlators, satisfy the topological recursion.
This applies in particular to all $(p,q)$ minimal models reductions of the KP hierarchy, or to the six Painlev\'e systems.
\end{center}
\begin{quote}

\end{quote}

\tableofcontents

\section{Introduction and setting}

First, we mention that this article is the generalization of \cite{P2,IwakiMarchal} (rank 2 systems) to arbitrary rank. The generalization is not straightforward and requires the new tools of \cite{BouchardEynard2016,Lie} since the loop equations and the spectral curves associated to arbitrary rank systems are far more involved.

\subsection{Generalities about quantum curves and mirror symmetry}

In the past few years, the interest has grown in the notion of ``{\bf quantum curves}'' related to enumerative geometry problems. In particular the relationship to the topological recursion \cite{EOFg} has raised a specific interest. Indeed, many enumerative geometry problems have two sides related by mirror symmetry (in fact they have a third side, namely integrability):
\begin{itemize}
\item a {\bf B model} side, described in terms of some algebraic manifold, typically a complex plane curve called the ``spectral curve'' and given by an algebraic equation:
\beq E_{\rm B}(x,y)=0 \eeq
Many invariants can be associated to a spectral curve, in particular the topological recursion invariants $W_{g,n}$ of \cite{EOFg}. Many recent papers have dealt with a ``quantization'' of that spectral curve, into a differential operator :
\beq
E_{\rm B}(x,y) \quad \overset{\text{quantization}}{\longrightarrow} \quad \hat E_{\rm B}(x,\hbar \frac{d}{dx},\hbar)
\eeq
such that $\hat E_{\rm B}(x,y,0)=E_{\rm B}(x,y)$ and such that it annihilates a ``wave function''
\beq
\hat E_{\rm B}(x,\hbar \frac{d}{dx},\hbar).\psi_{\rm B}(x,\hbar)=0
\eeq
In general, the wave function $\psi_{\rm B}(x,\hbar)$ has an $\hbar$ expansion of WKB type whose coefficients are certain combinations of the $W_{g,n}$'s associated to the spectral curve. In other words in the B-model side, the quantum curve $\hat E_{\rm B}(x,\hbar \frac{d}{dx},\hbar)$, as well as the wave function $\psi_{\rm B}(x,\hbar)$, are built from the classical spectral curve by the Topological Recursion.

\item an {\bf A model} side, describing an enumerative geometry problem, typically the enumeration of surfaces of given topologies together with some mapping into some target space. This includes many cohomological field theories, for example Gromov-Witten theory, as well as enumerations of maps, some conformal field theories, computing of knot invariants, random matrices...
In all these problems, there is a notion of enumerating geometric objects of a given ``genus'', and one can define some generating functions, with a formal parameter called $\hbar$ (rather denoted $g_s$ in topological string theory, or $1/N$ in random matrix theory), by summation over the genus. 
For example in Gromov-Witten theory, the generating function $W_{g,n}$ counts  the number of holomorphic maps of Riemann surfaces of genus $g$ with $n$ boundaries into a given Calabi-Yau manifold. The genus summation defines a formal series
\beq
W_n = \sum_{g=0}^\infty \hbar^{2g-2+n} W_{g,n}.
\eeq
All these formal generating functions $W_n$ can be put together to define a formal ``wave function'' $\psi_{\rm A}(x,\hbar)$ that encodes all the enumerative geometry.

\item {\bf Integrability}. In many such A-models, the geometry implies that the generating functions satisfy some equations (for instance gluing surfaces along their boundaries gives another surface), that can be encoded into an integrable system, such that the wave function $\psi_{\rm A}(x,\hbar)$ is its Baker-Akhiezer function. In other words, the geometric properties of the setup imply that the A-model's wave function $\psi_{\rm A}$ has to satisfy some differential system, again typically a quantum curve $\hat E_{\rm A}(x,\hbar \frac{d}{dx},\hbar).\psi_{\rm A}(x,\hbar)=0$.
For example the famous Witten-Kontsevich enumerative geometry problem of intersection theory on the moduli space of stable curves is related to the KdV integrable system. The corresponding wave function is simply the Airy function $\psi_{\rm A}(x,\hbar) = \text{Ai}(\hbar^{-2/3}x)$ and is annihilated by the operator $\hat E_{\rm A}=\hbar^2 \frac{d^2}{dx^2}-x$ which is a quantization of the classical spectral curve $E_{\rm A}(x,y)=y^2-x$. For cohomological field theories, the Dubrovin-Zhang \cite{DZ,DZ2} and the Givental \cite{G} formalisms also produce wave functions associated to integrable systems and quantum curves.

\item One of the main questions regarding mirror symmetry is then to prove that the A-model and B-model give rise to the same wave function  :
\beq
\psi_{\rm A}(x,\hbar) \overset{?}{=} \psi_{\rm B}(x,\hbar)
\quad , \quad
\hat E_{\rm A}(x,\hbar \frac{d}{dx},\hbar) \overset{?}{=} \hat E_{\rm B}(x,\hbar \frac{d}{dx},\hbar)
\eeq
In particular, since $E_B(x,y) =  \hat E_{\rm B}(x,y,0)$, it is easy to identify which B-model should be mirror to an A-model if we know $\hat E_{\rm A}$.
Notice that the equality holds in the sense of formal $\hbar$-series, so that we only need to work at the formal level.

\item In this article, we shall go from an A-model type integrable system towards a corresponding B-model. In other words, we start from a wave function annihilated by a differential operator in some integrable hierarchy, and prove, under certain assumptions on the differential operators, that its WKB expansion defines some $W_{g,n}$ differentials that obey the B-model topological recursion.
\end{itemize}

\bigskip
Our method is a generalization to systems of arbitrary rank of what was done in \cite{IwakiMarchal,P2} for $2\times 2$ systems. In \cite{BBE14,Lie}, it was proved that if a differential system satisfies the so-called {\bf Topological Type (TT)} property, then the corresponding $W_{g,n}$'s necessarily satisfy the topological recursion. 

What we do in this article is therefore to prove that a large class of integrable systems do satisfy the TT property.

\medskip

We mention that there exist many other articles \cite{BouchardEynard2016,MS12,MS15,N15,Iwaki,Penkava,Do,DM14,Quantum,QuantumDo}, taking (case by case) the opposite path. Starting from a B-model (i.e. a classical spectral curve), they construct $\psi_{\rm B}$ by the topological recursion and prove that there exists a quantum curve of isomonodromic type (and thus related to an integrable system) that annihilates $\psi_{\rm B}$.
 
Until today, there is no general theorem stating what kind of B-model spectral curve leads to an isomonodromic differential system (a quantum curve), and vice versa, there is no general theorem stating what kind of isomonodromic differential system has a WKB expansion governed by topological recursion. At the moment, all existing articles prove a correspondence within some specific subclasses. Most studied examples are rank 2 systems that are easier to study.

This article pursues a similar goal, extending a known proof for certain rank 2 systems to higher rank systems. It provides sufficient conditions for a differential system to have a WKB expansion governed by the topological recursion. The sufficient conditions are general enough so that they may be applied to many differential systems. In particular, they allow to recover all known cases like \cite{BE09,BBE14,P5,P2,IwakiMarchal,N15,Iwaki,Do}.

\bigskip

{\bf Outline :}
\begin{itemize}
\item We first introduce the two compatible differential systems, the corresponding Lax equations and some useful notations.
\item In section \ref{S2}, we state the assumptions required for our result. They rely on describing the algebro-geometric properties of the two underlying spectral curves --the eigenvalue loci of the Lax pair elements. Our assumptions are then that the main spectral curve has genus $0$, that the auxiliary one is an embedding without self-intersections, and the requirement that subleading terms are less singular than leading ones in the $\hbar$-expansions. These assumptions are indeed satisfied for many famous integrable systems.
\item In section \ref{SectionM}, we recall the definitions of correlation functions by determinantal formulas and of their formal WKB $\hbar$-expansion.
\item In section \ref{S4}, we prove our main result : the correlators built from a Lax pair satisfying our assumptions, are of ``topological type'' (we recall the definition), which in turn implies that they satisfy topological recursion.
\item Section \ref{S5} is our summary and conclusion. We mention what generalizations we may expect for non-zero genus spectral curves as well as the issue of the WKB expansion of the wave function in Appendix \ref{appequWKBPsiM}.
\item In Appendix \ref{AppendixExamples}, we show examples of classical integrable systems that satisfy our assumptions.
\end{itemize}

\subsection{Setting: compatible linear differential systems}

Let $\hbar>0$ be given (usually called ``formal expansion parameter'', ``dispersion parameter'', ``Planck constant'' or just ``small parameter''). As in \cite{Lie}, the natural context is the one of a reductive complex Lie algebra $\Lieg$ and its associated connected Lie group $G=e^{\Lieg}$, here we will however mostly restrict ourselves to the case $G=GL_d(\mathbb C)$ and $\Lieg=\mathfrak{gl}_d(\mathbb C)$ (this is the most common setting in practice) and leave the study with general $\Lieg$ for a later work currently under preparation \cite{ToCome}. 

Instead of a linear differential operator $\hat E(x,\hbar \frac{d}{dx}, \hbar)$ of order $d$ acting on a scalar wave function $\psi(x,\hbar)$, we consider an equivalent (and in fact more general) order one linear matrix differential system. More precisely we consider a time-dependent family of such systems:

\begin{itemize}

\item
We shall consider a compatible system of linear equations of the form :
\bea \label{GeneralLax} 
\hbar \partial_x \Psi(x,t,\hbar)&=& L(x,t,\hbar)\Psi(x,t,\hbar)\cr
\hbar \partial_t \Psi(x,t,\hbar)&=& R(x,t,\hbar)\Psi(x,t,\hbar)
\eea
where the  $d\times d$ matrix $\Psi(x,t,\hbar)\in GL_d(\mathbb C)$ is assumed invertible for all $x$ (in the general setting : $\Psi(x,t,\hbar)\in G$).
The $d\times d$ matrices $L(x,t,\hbar)$ and $R(x,t,\hbar)$ (in the general setting $L,R \in\Lieg$) are assumed to be \textbf{rational functions} of $x$ for any values of $t$ and $\hbar$. $x$ is usually called a ``spectral parameter'' and $t$ a ``time parameter''. Note that to shorten notations, \textbf{we shall often write only the $x$ dependence explicitly and drop the $t$ or $\hbar$ dependence in the notations}:
\beq
L(x,t,\hbar) \stackrel{{\rm notation}}{\equiv} L(x), \dots
\eeq

We shall prove in this article that a set of conditions on $L$ and $R$ are sufficient for the system to be of ``Topological Type''. 

\item The compatibility relation of the two equations is called the Lax equation :
\beq \label{Laxeq1} 
\hbar \partial_t L(x,t,\hbar) - \hbar \partial_x R(x,t,\hbar) = [R(x,t,\hbar),L(x,t,\hbar)].
\eeq

\item \underline{Generalization to arbitrary $G$} : The system of equations \eqref{GeneralLax} can be viewed, with $x\in \curve_0$ in  a local coordinate patch on a complex curve, as the equation defining a flat section $\Psi(x,\hbar)\in G$ of a stable principal $G$--bundle $\mathcal E=G\to \curve_0$ -- with $G$ a connected reductive Lie group -- over $\curve_0$, equipped with a meromorphic connection $\nabla=d - \hbar^{-1}  L(x,\hbar)dx$, where $L(x,\hbar)dx$ is a $\Lieg$--valued meromorphic one form on $\curve_0$. In this general context, $x$ is called a spectral parameter, $\hbar^{-1}L(x,\hbar)dx$ is called a Higgs field, and the pair $(\mathcal E,L)$ is called a Hitchin pair.
Here we shall restrict ourselves to the Riemann sphere $\Sigma_0=\bar{\mathbb C}=\mathbb C\cup\{\infty\} $.

\item In the construction of \cite{BE09,BBE14}, to a solution $\Psi(x,t,\hbar)\in G$ of the differential system is associated a solution $M(x,t,\hbar)\in \Lieg$ of the adjoint system :
\bea\label{Msys}
\hbar \partial_x M(x,t,\hbar) &=& [L(x,t,\hbar),M(x,t,\hbar)]\cr
\hbar \partial_t M(x,t,\hbar) &=& [R(x,t,\hbar),M(x,t,\hbar)]
\eea 
whose solutions are of the form
\beq
\Psi(x,t,\hbar) E \Psi(x,t,\hbar)^{-1}
\eeq
where $E$ is a constant (in the sense $\partial_x E=0$) element of $\Lieg$. We will therefore denote them as :
\beq
M(x.E) \overset{\text{notation}}{{\equiv}} M(x.E,t,\hbar) = \Psi(x,t,\hbar) E \Psi(x,t,\hbar)^{-1},
\eeq
often not writing the $t$ and $\hbar$ dependence to lighten notations.
Notice that any another solution of \eqref{GeneralLax} is obtained from $\Psi$ through the right multiplication $\Psi(x)\to \Psi(x) C$ by a constant matrix $C\in G$, $\partial_x C=0$. $M(x.E)$ then changes to $M(x.CEC^{-1})$, i.e. an adjoint transformation of $E$. Note also that $M(x.E)$ depends linearly on $E\in\Lieg$.
Since $\Psi$ has monodromies around the singularities of $L(x)$, $x$ lives on the universal cover $\tilde \Sigma_0$ of $\Sigma_0\setminus \{{\rm singularities\,of\,} L \}$. 
\item Then, still following \cite{BE09,BBE14}, one associates to the differential system of the form \eqref{GeneralLax} or \eqref{Msys}, a sequence of local $n$-forms $(\omega_n)_{n\geq 1}$ on $(\tilde\curve_0\times \Lieg)^n$ usually called the ``correlators'' : 
\bea \label{DefWn}
\omega_n(x_1.E_1,\dots,x_n.E_n)
= 
\left\{
\begin{array}{lr}
\hbar^{-1} \operatorname{Tr} \left(L(x_1)M(x_1.E_1)\right)dx_1 & n=1 \cr
\frac{1}{n}\underset{\sigma\in\mathfrak S_n}{\sum} \frac{\operatorname{Tr} \underset{i=1}{\overset{n}{\prod}} M(x_{\sigma(i)}.E_{\sigma(i)}) }{\underset{i=1}{\overset{n}{\prod}} (x_{\sigma(i)}-x_{\sigma(i+1)})} \underset{i=1}{\overset{n}{\prod}} dx_i& n\geq 2
\end{array}\right.
\eea
or equivalently a sequence of functions $(W_n)_{n\geq 1}$:
\bea \label{DefWn2}
W_n(x_1.E_1,\dots,x_n.E_n)
= 
\left\{
\begin{array}{lr}
\hbar^{-1} \operatorname{Tr} \left(L(x_1)M(x_1.E_1)\right) & n=1 \cr
\frac{1}{n}\underset{\sigma\in\mathfrak S_n}{\sum} \frac{\operatorname{Tr} \underset{i=1}{\overset{n}{\prod}} M(x_{\sigma(i)}.E_{\sigma(i)}) }{\underset{i=1}{\overset{n}{\prod}} (x_{\sigma(i)}-x_{\sigma(i+1)})} & n\geq 2
\end{array}\right.
\eea
where a choice is made once and for all for writing non-commutative products.
In the generalization to a Riemann surface $\Sigma_0$ instead of $\bar{\mathbb C}$, the $\frac{\sqrt{dx_i dx_j}}{(x_i-x_j)}$ terms are replaced by inverses of twisted Fay's prime forms.
In $G=GL_d(\mathbb C)$ the trace of a product is defined in the defining representation (i.e. the usual trace for $d\times d$ matrices). For other Lie groups, we define the trace by choosing the matrix-trace in a once for all given faithful representation. These correlators $\omega_n$ and $W_n$ appear naturally in matrix models and in many enumerative problems \cite{BE09,BBE14,Lie,BookEynard,BDY1,BDY2}.

\item In \cite{Lie}, it was proved that these $W_n$'s always satisfy a family of equations called ``{\bf loop equations}'' (and that are analogous to Virasoro or W-algebra constraints, in the CFT context, see \cite{MS15}). This is important because loop equations can be solved recursively in terms of $\hbar$-expansions.

\item The WKB asymptotics are defined as formal $\hbar$-series solutions to \eqref{GeneralLax} with $\hbar$ assumed to be small. They take the form :
\beq
\Psi(x,t,\hbar) \mathop{\sim}_{\hbar\to 0} V(x,t) \left( {\mathbf 1}_d + \underset{k=1}{\overset{\infty}{\sum}} \hbar^k \hat\Psi^{(k)}(x,t) \right) e^{\frac{1}{\hbar} T(x,t)} C
\eeq
that we shall explain below in section \ref{SectionM}, let us just mention here that $T(x,t)$ is diagonal (and more generally belongs to a Cartan subalgebra $\Lieh\subset \Lieg$).
In turn, this implies that if $CEC^{-1}$ is diagonal ($\operatorname{Adj}_C E\in\Lieh$), then $M(x.E,t,\hbar)$ has a formal $\hbar$-series expansion of Taylor type (which is a special case of formal $\hbar$-WKB expansion but without an exponential factor):
\beq
M(x.E,t,\hbar) = \underset{k=0}{\overset{\infty}{\sum}} \hbar^k {\check M}^{(k)}(x.E,t),
\eeq
Consequently all correlators $W_n$ defined by \eqref{DefWn} also admit a $\hbar$-formal Taylor expansion that we shall denote:
\bea
\omega_n(x_1.E_1,\dots,x_n.E_n) &=& \underset{k=-\delta_{n,1}}{\overset{\infty}{\sum}} \hbar^{k} \omega_n^{(k)}(x_1.E_1,\dots,x_n.E_n)\cr
W_n(x_1.E_1,\dots,x_n.E_n) &=& \underset{k=-\delta_{n,1}}{\overset{\infty}{\sum}} \hbar^{k} W_n^{(k)}(x_1.E_1,\dots,x_n.E_n).
\eea
\textbf{The main questions are then the following :
\begin{enumerate}\item Is there a general method to compute the coefficients $W_n^{(k)}$?
\item Taking $k=2g-2+n$, does $\omega_n^{(k)}$ coincide with $\omega_{n,g}$ computed from the topological recursion? 
\end{enumerate}
}

\item In \cite{BBE14,Lie} some sufficient conditions on the differential systems, known as the ``Topological Type'' (TT) property, were given to get a positive answer. The goal of this article is to find a large class of Lax pairs $(L,R)$ satisfying the TT property.

\end{itemize}

\section{Assumptions \label{S2}}

We shall now describe our assumptions on the pair $(L(x,t,\hbar),R(x,t,\hbar))$. These assumptions are described in terms of algebraic geometry and the notion of spectral curve. All in all, there are $6$ assumptions that are presented in the following subsections. Each assumption allows new definitions and/or implies new properties that are presented in several lemmas and propositions. Although they may appear technical, these assumptions have been proved to hold in many cases like Painlev\'{e} Lax pairs \cite{P5,P2,IwakiMarchal}. We also prove them for all $(p,q)$ minimal models in Appendix \ref{AppendixExamples}.  

\subsection{Spectral curve(s)}

\begin{assumption}[$\hbar$ expansion]\label{Asshbarexpansion}
We make the assumption that $L(x,t,\hbar)$ and $R(x,t,\hbar)$ have a limit at $\hbar\to 0$ :
\beq
\lim_{\hbar\to 0} L(x,t,\hbar) = L^{(0)}(x,t)
\quad , \quad
\lim_{\hbar\to 0} R(x,t,\hbar) = R^{(0)}(x,t),
\eeq
and that both limits are rational functions of $x$. Furthermore, we assume that $L(x,t,\hbar)$ and $R(x,t,\hbar)$ have an $\hbar$ expansion (formal or asymptotic) of the form :
\beq
L(x,t,\hbar) = \sum_{k=0}^\infty \hbar^k\,L^{(k)}(x,t)
\quad , \quad
R(x,t,\hbar) = \sum_{k=0}^\infty \hbar^k\,R^{(k)}(x,t).
\eeq
where all $L^{(k)}(x,t)$ and $R^{(k)}(x,t)$ with $k\geq 0$ are rational functions of $x$.
\end{assumption}

The spectral curve is defined as the zero locus of the characteristic polynomial of the matrix $L^{(0)}$, i.e. the eigenvalues locus, whence the name ``spectral" curve. In the general Lie group context, this corresponds to the Hitchin's map.

\begin{definition}[Spectral curve] The (family of) spectral curve of the differential system is the zero locus of the characteristic polynomial in $\mathbb C\times \mathbb C$ :
\beq \label{SpecCurve} 
\spcurve \equiv \spcurve_t=
\{(x,y)\in \mathbb{C}^2 \text{ such that } E_t(x,y)=\det(y-L^{(0)}(x,t))=0\}
\eeq
This defines an (a family of) algebraic plane curve immersed into $\mathbb C\times \mathbb C$. We define the two meromorphic functions corresponding to the $x$ and $y$ projection in $\mathbb C\times \mathbb C$:
\bea
\x\equiv \x_t :\spcurve_t \to \mathbb C
\quad , \quad 
\y\equiv \y_t :\spcurve_t \to \mathbb C \\
(x,y)\mapsto x
\quad \, \quad 
(x,y)\mapsto y.
\eea
The plane curve can be desingularized. Its desingularization is a smooth compact Riemann surface noted $\curve\equiv \curve_t$, and the functions $\x\equiv \x_t$ and $\y\equiv \y_t$ can be identified with meromorphic functions $\curve_t\to \mathbb C$. This allows to redefine the (family of) spectral curve as the triple:
\beq \label{SPCurve}
\spcurve\equiv \spcurve_t=(\curve_t,\x_t,\y_t),
\eeq
given by a (family of) compact Riemann surface $\curve\equiv \curve_t$, equipped with two meromorphic functions $\x_t :\curve_t\to \mathbb C$ and $\y_t :\curve_t\to \mathbb C$. On a compact curve, any two meromorphic functions are related by an algebraic equation:
\beq
\forall\,z\in \curve_t\quad ,\quad
E_t(\x_t(z),\y_t(z))=0 \,\,\, \text{  where } E\equiv E_t \text{ is a (family of) polynomial}
\eeq
thus giving an alternative definition of the spectral curve directly from \eqref{SPCurve}.
We shall also be interested in the (family of) one-form $\omega_1^{(0)}$ on $\curve_t$ defined by $\omega_1^{(0)}=\y d\x$, sometimes called the Liouville form, because it is the pullback to the spectral curve, of the tautological form of $\mathbb C\times \mathbb C$, viewed as the cotangent space of $\mathbb C$.

The $y$--degree of the characteristic polynomial is the matrix's size (the rank of $GL_d$):
\beq
\deg_y E_t(x,y) =d
\eeq
and thus for a given generic $x\in \mathbb C$, the equation $E_t(x,y)=0$ has $d$ solutions, that are the $d$ eigenvalues $Y_1(x,t),\dots,Y_d(x,t)$ of $L^{(0)}(x,t)$. They are the images by the function $\y_t$, of the $d$ preimages of $x$  by $\x_t$ :
\beq
\x_t^{-1}(x) =\{z\in\curve_t \,|, \x_t(z)=x\} = \{z^1(x,t),\dots,z^d(x,t)\}
\eeq
gives
\beq
Y_i(x,t)=\y_t(z^i(x)).
\eeq
Here the ordering of eigenvalues is arbitrary and can always be locally chosen analytical within some open simply connected domain of $\mathbb{C}\setminus \{\text{Singularities of } \x_t \text{ and } \x_t^{-1} \}$. The ordering will turn out to be irrelevant for our purposes, and we assume it chosen within such domain once and for all.

\end{definition}

\begin{definition}[Auxiliary spectral curve]
In the same spirit we define the (family of) auxiliary spectral curve by the zero locus of the characteristic polynomial of $R^{(0)}$ :
\beq \label{AuxCurve} 
\tilde \spcurve\equiv \tilde \spcurve_t =
\{(x,s)\in \mathbb{C}^2 \text{ such that } \td{E}_t(x,s)=\det(s-R^{(0)}(x,t))=0\}
\eeq
that we shall encode as the triple
\beq
{\tilde\spcurve}_t=(\tilde \curve_t,\tilde \x_t,\s_t),
\eeq
given by a (family of) Riemann surface $\tilde\curve_t$, equipped with two meromorphic functions $\tilde \x_t :\tilde\curve_t\to \mathbb C$ and $\s_t :\tilde\curve_t\to \mathbb C$, related by the algebraic equation
\beq
\forall\,z\in \tilde\curve_t\quad ,\quad
\tilde E_t(\tilde \x_t(z),\s_t(z))=0.
\eeq
Similarly for a given $x$, there exist $d$ solutions noted $(S_1(x,t),\dots,S_d(x,t))$ of the auxiliary curve $\td{E}_t(x,s)=0$. They are the $d$ eigenvalues of $R^{(0)}(x,t)$, and also $S_i(x,t) = \s_t(\tilde z^i_t(x))$ with $\tilde \x_t(\tilde z^i_t(x))=x$.
\end{definition}

\bl
The matrices $L^{(0)}(x,t)$ and $R^{(0)}(x,t)$ commute thus they generically have a common basis of eigenvectors and their eigenvalues are not algebraically independent. In particular the spectral curves $\spcurve_t$ and $\tilde\spcurve_t$ have the same desingularization : $\curve_t=\tilde\curve_t$ and the same $\x$-projection to $\bar{\mathbb C}$ : $\tilde \x_t=\x_t$.

\el

\proof{
At order $\hbar^0$, the Lax compatibility condition \eqref{GeneralLax} reads:
\beq
[L^{(0)}(x,t),R^{(0)}(x,t)]=0.
\eeq
For generic $x$, all the eigenvalues of $R^{(0)}(x,t)$ are distinct. It implies that the set of matrices commuting with $R^{(0)}(x,t)$ is the algebra of polynomials of $R^{(0)}(x,t)$.
Consequently there exists a polynomial $Q(x,s)$ (the interpolating Lagrange polynomial) such that $L^{(0)}(x,t) = Q(x,R^{(0)}(x,t))$, and $Y_i(x)=Q(x,S_i(x))$, i.e. $\y(z)=Q(\tilde \x(z),\s(z))$ for all $z\in \tilde \curve_t$. This implies that $\y_t$ is a meromorphic function on $\tilde\curve_t$. Permuting the roles of $R^{(0)}$ and $L^{(0)}$ also shows that $\s_t$ is a meromorphic function on $\curve_t$. Therefore $\curve_t=\tilde\curve_t$, and $\x_t=\tilde \x_t$.
}

\subsection{Geometry of the spectral curve}

\subsubsection{Genus $0$ assumption}

From now on, we shall assume that our system is such that :

\begin{assumption}[Genus zero Spectral Curve]\label{genuszeroassumption}
The compact Riemann surface $\curve_t$ has genus equal to $0$. This implies that it is isomorphic to the Riemann sphere $\curve_t=\bar{\mathbb C}=\mathbb C\cup\{\infty\}=\mathbb CP^1$  (the complex plane compactified at $\infty$), and that, for any given $t$ in an open domain, the functions $\x_t,\y_t,\s_t$, are rational functions of a variable $z\in\bar{\mathbb C}$ :
\beq
\x_t(z),\y_t(z),\s_t(z)\,\, \in \mathbb C(z)=\{ {\rm \,\,rational\,\,functions\,\,of}\,\,z \}
\eeq

\end{assumption}

\br
The issue of determining if this genus zero hypothesis can be lifted is mostly open. In fact in the example of matrix models, it is known that the TT property is generically not satisfied when the genus is strictly positive. But a generalization of the TT property can be found by allowing the coefficients in the $\hbar$ expansion, to be ``oscillatory'', i.e. bounded quasi-periodic functions of $\frac{1}{\hbar}$. In that case, the oscillatory terms are themselves found by the topological recursion (See \cite{EynMath}).
 
Besides, in knot theory, the TT property happens to hold with spectral curves (A-polynomial) of strictly positive genus. This is due to a miracle that the $\frac{1}{\hbar}$ term is exactly a period of the oscillatory term, and thus can be treated as a constant coefficient, see \cite{BEknots}, and then the TT property holds.
So the general situation is still unclear.
\er

\br
The choice of the parameterizing variable $z$ is arbitrary up to M\"obius transformations (automorphisms of the Riemann sphere) :
\beq
z\mapsto \frac{az+b}{cz+d}.
\eeq
In particular, we may chose the coefficients of the M\"obius transformation $a,b,c,d$  to be time dependent : $a(t),b(t),c(t),d(t)$.
\er

The functions $\x_t,\y_t,\s_t$ are functions of both $z$ and $t$, and they are rational in $z$.
We mention that their dependence on $t$ does not need to be rational. In many examples they are transcendental functions of $t$, like for example solutions of Painlev\'e equations.

We shall denote for any function $f(z,t)$ :
\beq\label{notationprimedot}
f'(z,t) = \frac{\partial f}{\partial z}
\qquad , \qquad
\dot f(z,t) = \frac{\partial f}{\partial t}.
\eeq
Note that taking a time derivative at fixed $x(z,t)$, following from the chain rule, takes the form of a Poisson bracket $\{f,\x\}=\dot f \, \x' - \dot \x\, f'$ :
\beq
\left.\frac{d f(z,t)}{dt}\right|_{x(z,t)} = \dot f - \dot \x\,\frac{f'}{\x'}
= \frac{\dot f \, \x' - \dot \x\, f'}{\x'} = \frac{1}{\x'}\, \{f,\x\},
\eeq
thus reflecting the symplectic structure of $\mathbb C \times \mathbb C$ of which the family $\{\Sigma_t\}_t$ defines a Lagrangian foliation.

\subsubsection{Behavior at poles}

\bl
The poles of the eigenvalues $\y_t(z)$ (resp. $\s_t(z)$) are poles of $L^{(0)}(\x(z),t)$ (resp. $R^{(0)}(\x(z),t)$) of at least the same order.
\el

\proof{
Let $\alpha$ be a pole of $\y_t(z)$ of order $d_\alpha>0$ so that $\y_t(z)=O\left((z-\alpha)^{-d_\alpha}\right)$.Let us assume that  $(z-\alpha)^{d_\alpha} L^{(0)}(\x(z),t)=o(1)$. This would imply that
\beq
0=\det(\y_t(z){\rm Id}-L^{(0)}(\x(z),t) ) =\y_t(z)^d (1+o(1))
\eeq
which is a contradiction. This implies that $L^{(0)}(\x(z),t)$ has a pole of order at least $d_\alpha$. Obviously, the same holds for $R^{(0)}(x,t)$.
}

\bl[Poisson relation]
The eigenvalues $\left(Y_i(x,t)\right)_{1\leq i\leq d}$ of $L^{(0)}(x,t)$ and $\left(S_i(x,t)\right)_{1\leq i\leq d}$ of $R^{(0)}(x,t)$ are related by :
\beq
\frac{\partial Y_i(x,t)}{\partial t} = \frac{\partial S_i(x,t)}{\partial x}.
\eeq
Equivalently, the functions $\x_t(z),\y_t(z),\s_t(z)$ satisfy :
\beqq
\frac{\partial \y_t(z)}{\partial t}\, \frac{\partial \x_t(z)}{\partial z}
- \frac{\partial \x_t(z)}{\partial t}\, \frac{\partial \y_t(z)}{\partial z}
=  \frac{\partial \s_t(z)}{\partial z},
\eeqq
or written in the notations of \eqref{notationprimedot} :
\beqq
\{\y_t,\x_t\} = \dot \y_t \x_t' - \dot \x_t \y_t' =\s_t'.
\eeqq
\el

\proof{
Since $L^{(0)}(x,t)$ and $R^{(0)}(x,t)$ commute, they generically have a common basis of eigenvectors, let us denote $V(x,t)$ the matrix whose $i^{\rm th}$ column is the eigenvector of $L^{(0)}(x,t)$ with eigenvalue $Y_i(x,t)$ and of $R^{(0)}(x,t)$ with eigenvalue $S_i(x,t)$.
Denoting $Y(x,t)={\rm diag}(Y_1(x,t),\dots,Y_d(x,t))$ and $S(x,t)={\rm diag}(S_1(x,t),\dots,S_d(x,t))$, we have
\beq
L^{(0)}(x,t)=V(x,t) Y(x,t) V(x,t)^{-1}
\quad , \quad
R^{(0)}(x,t)=V(x,t) S(x,t) V(x,t)^{-1}.
\eeq
Now write the Lax equation to order $\hbar^1$ and conjugate by $V(x,t)$ :
\bea
[S(x,t),V(x,t)^{-1} L^{(1)}(x,t)V(x,t)]+[V(x,t)^{-1}R^{(1)}(x,t)V(x,t),Y(x,t)] \cr
= \partial_t Y(x,t) - \partial_x S(x,t) 
\eea
The left hand side is a sum of commutators with diagonal matrices, hence has vanishing entries on the diagonal. On the contrary, the right hand side is a diagonal matrix and evaluating its diagonal entries gives the sought result :
\beq
0 = \partial_t Y(x,t) - \partial_x S(x,t).
\eeq
}

\medskip

As an immediate corollary we get :
\bc
Finite (i.e at $x\neq \infty$) singularities of $S$ are also singularities of $Y$, of at least the same degree.
And if $S$ has a singularity of order $d_\infty$ at $x=\infty$, then $Y$ has a singularity at $x=\infty$ of order at least $d_\infty+1$.
\ec

Note that the converse is not true : some singularities of $Y$ may be time independent and may not be singularities of $S$. In some sense, we can say that $R^{(0)}$ is less singular than $L^{(0)}$.

\subsubsection{Branchpoints and double points}

\bd[Branchpoints]
We define the branchpoints $\left(a_i\right)_{1\leq i\leq r}$ as the points of $\curve$ where the map $z\mapsto \x(z)$ is not locally invertible. There may be two kinds of branchpoints :

\begin{itemize}

\item Finite branchpoints, at which $\x(a_i)\neq \infty$. They are zeros of the differential $d\x$ :
\beqq
d\x(a_i)=0.
\eeqq 
Moreover, they are among the simultaneous solutions of $E(x,y)=0$ and $E_y(x,y)\equiv\partial_y E(x,y)=0$.

\item Branchpoints at poles of $\x$ of order $\geq 2$. 

\end{itemize}

A branchpoint $a_i$ of the spectral curve $\spcurve$ (resp. $\tilde \spcurve$) is called \textit{regular} if it is not a branchpoint of $\y$ (resp. $\s$).
Generic finite branchpoints of $\x$ have order 2, i.e. are simple zeros of $d\x$, and regularity means that they are not zeros of $d\y$ (resp. $d\s$).
\ed

Note that the branchpoints may depend on time $t$. However, the number of branchpoints $r\geq 1$ does not locally depend on $t$. We will also need the following definition:

\bd[Double points (also called self-intersections)]
We define the double points $(\, (b_i,\bar b_i)\, )_{1\leq i\leq r''}$ (resp. $( (\tilde b_i, \bar{\tilde b}_i) )_{1\leq i\leq \tilde r''}$) of the curve $\spcurve=(\curve,\x,\y)$ (resp. of $\tilde\spcurve=(\curve,\x,\s)$), as the pairs $(b_i,\bar b_i)=(z,z')$ (resp. $(\tilde b_i,\bar{\tilde b}_i)=(z,z')$) solutions of
\beq
\left\{
    \begin{array}{rcl}
        \x(z) & = & \x(z') \\
        \y(z) & = & \y(z') \\
        z&\neq& z'
    \end{array}
\right.
\qquad , \qquad 
\left( {\rm resp.}\quad
\left\{
    \begin{array}{rcl}
        \x(z) & = &\x(z') \\
        \s(z) & = &\s(z') \\
        z&\neq& z'
    \end{array}
\right.
\right)
\eeq
These double points $(x,y)=(\x(b_i),\y(b_i))=(\x(\bar b_i),\y(\bar b_i))\in \mathbb C\times \mathbb C$ of the spectral curve (resp. $(x,s)=(\x(b_i),\s(b_i))=(\x(\bar b_i),\s(\bar b_i))\in \mathbb C\times \mathbb C$), are then solutions of the system
\beq
\left\{
    \begin{array}{rcl}
E(x,y)&=&0 \\
E_y(x,y)&=&0 \\
E_x(x,y)&=&0
    \end{array}
\right.
\qquad , \qquad 
\left( {\rm resp.}\quad
\left\{
    \begin{array}{rcl}
\tilde E(x,s)&=&0 \\
\tilde E_s(x,s)&=&0 \\
\tilde E_x(x,s)&=&0
    \end{array}
\right.
\right)
\eeq
\ed

We shall make the following assumption regarding the double points of the auxiliary spectral curve :

\begin{assumption}[Regularity of $\spcurve_t$ and no double points for $\td{\spcurve}_t$]\label{Assnodblpt}
We make the assumption that the auxiliary spectral curve $\tilde\spcurve_t$ is regular and has no double points. In other words, $\tilde\spcurve_t$ is a smooth embedding into $\mathbb C \times \mathbb C$ (rather than an immersion) with no self-intersection. Moreover we assume that $\spcurve_t$ is regular.
\end{assumption}

Note that the last assumption does not exclude the possibility that the spectral curve $\spcurve_t$ admits double points. Moreover, the auxiliary spectral curves $\tilde \spcurve_t=(\bar{\mathbb C},\x_t,s_t)$ satisfying assumptions \ref{genuszeroassumption} and \ref{Assnodblpt} are the same as the ones described in \cite{BouchardEynard2016}.

\medskip

We have the following lemma :

\bl\label{lemmadxsurEyhol}
The meromorphic one-form
\beqq
\frac{d\x(z)}{E_y(\x(z),\y(z))}
\eeqq
is holomorphic at all branchpoints (finite or infinite). It has poles only at double points (generically simple poles at $b_i$ and $\bar b_i$ with opposite residues) and/or at simple poles of $\x$.
\el

\proof{
This is a classical algebro-geometric result, we refer to \cite{Fay}. Let us sketch the proof. Near a finite branchpoint $a$ of order $k\geq 2$, $z=(x-\x(a))^{1/k}$ can be used as a local coordinate. Consider the case $\y(a)\neq \infty$. Since the branchpoint is regular, $d\y$ does not vanish at that point, i.e.
\beq
\y(z)= \y_0 + \y_1 z +O(z^2), \qquad \y_1\neq 0.
\eeq
This gives :
\beq
E(x,y)= ((y-\y_0)^k-\y_1^k (x-\x(a)) ) \times (1+o(1)),
\eeq
and
\beq
\frac{d\x}{E_y(\x,\y)} = \frac{k z^{k-1} dz}{k (\y-\y_0)^{k-1}} \times (1+o(1))
\eeq
and thus $\frac{d\x}{E_y(\x,\y)}$ is analytic at $z=0$, i.e. at $x=a$.
The other cases where $\x(a)=\infty$ or $\y(a)=\infty$ can be treated similarly in a local variable. In other words, for finite regular branchpoints, both $d\x(z)$ and $E_y(\x(z),\y(z))$ vanish, at the same order so that the ratio remains finite.

For double points, $E_y(\x(z),\y(z))$ vanishes but not $d \x(z)$, so that the ratio has a pole. Writing
\beq
E(x,y) = \prod_{k=1}^d (y-\y(z^k(x)))
\eeq
we have that when $z\to b_i$, simultaneously $z'\to \bar b_i$, and
\beq
E_y(\x(z),\y(z)) \sim (\y(z)-\y(z')) E_{y,y}(\x(b_i),\y(b_i))
\sim (z-z')\,\frac{d\y(z)}{dz}\, E_{y,y}(\x(b_i),\y(b_i)).
\eeq
Assuming that the double point is generic, i.e. $E_{y,y}d\y\neq 0$, we get :
\beq
\Res_{z\to b_i} \frac{d\x(z)}{E_y(\x(z),\y(z))} 
= - \Res_{z\to \bar b_i} \frac{d\x(z)}{E_y(\x(z),\y(z))} 
= \frac{d\x(b_i)}{d\y(b_i) \,E_{y,y}(\x(b_i),\y(b_i))} 
\eeq
}

\subsection{Eigenvectors}

Let ${\rm div}_\infty \x = \underset{k=1}{\overset{p}{\sum}} d_k \alpha_k$ be the divisor of poles of the rational function $z\mapsto \x(z)$, with $d_k$ the degree of $\alpha_k$ ($\alpha_k$ may depend on $t$). The total degree is the size of the matrix
\beq
\underset{k=1}{\overset{p}{\sum}} d_k=d.
\eeq
Up to a M\"obius change of variable on $z$, we may assume that none of the $\alpha_k$'s is at $\infty$. We can thus can write the rational function $\x(z)$ uniquely as :
\beq\label{xsumpoles}
\x(z) = X_{\infty,0} + \underset{k=1}{\overset{p}{\sum}}\sum_{l=1}^{d_k} \frac{X_{k,l}}{(z-\alpha_k)^{l}}.
\eeq
where $d_k\geq 1$. Moreover, if $d_k\geq 2$ then $\alpha_k$ is a (non-finite) branchpoint. 
Note that if one of the $\alpha_k$ is at $\alpha_\infty=\infty$ we would rather write :
\beq
\x(z) = \sum_{l=0}^{d_\infty} X_{\infty,l} z^l  + \underset{k=1}{\overset{p}{\sum}}\sum_{l=1}^{d_k} \frac{X_{k,l}}{(z-\alpha_k)^{l}},
\eeq
But to avoid useless notation complications, upon changing $z$ by a M\"obius transformation, we shall assume that all poles of $\x(z)$ are different from $\infty$.

\subsubsection{The generalized Vandermonde matrix $\mathcal V(x)$}

\bd\label{defV}
For generic points $z\in\Sigma$, in particular away from the branchpoints, let us define the $d$-dimensional vector $\vec{\mathcal V}(z)$ with entries labeled by all possible pairs $(k,l)$ with $1\leq k\leq p$ and $1\leq l\leq d_k$ :
\beq
\vec{\mathcal V}(z) = (\mathcal V_{k,l}(z))_{k,l}
\quad , \quad
{\rm where}\,\,
\mathcal V_{k,l}(z) = \frac{1}{(z-\alpha_k)^{l}\sqrt{\x'(z)}}.
\eeq
In addition we define these matrix entries to be ordered as follows
\beq
\vec{\mathcal V}(z)=\left(\mathcal{V}_{1,1}(z),\dots,\mathcal{V}_{1,d_1}(z),\dots,\mathcal{V}_{p,1}(z),\dots,\mathcal{V}_{p,d_p}(z)\right).
\eeq 

Let $\mathcal V(x)$ be the $d\times d$ square matrix whose columns are the vectors $\vec{\mathcal V}(z^j(x))$ :
\beq
\forall\, 1\leq k\leq p, 1\leq l\leq d_k, 1\leq j\leq d\, :\,\, (\mathcal V(x))_{k,l;j} = \mathcal{V}_{k,l}(z^j(x))
\eeq
It is analytic locally in some open simply connected domain, in which the $z^i$ and the square root are defined.
\ed

\br
The sign of the square root, chosen arbitrarily, is well defined and locally analytic within some open simply connected domain -- the same domain in which we defined the ordering of $z^i(x)$.
In fact in all what follows, the square root will almost always appear to the power two, so the sign will eventually be irrelevant.
\er

\br
Note that if $\x$ has only one pole ($p=1$ and $d_1=d$) then the previous matrix is a Vandermonde matrix multiplied by $\frac{1}{(z-\alpha_1)\sqrt{\x'(z)}}$, hence the name ``generalized Vandermonde matrix''.
\er

The matrix $\mathcal{V}(x)$ satisfies remarkable properties.

\bl\label{lemmaVtCV}
There exists an invertible $d\times d$ matrix $C\equiv C(t)$ (independent of $x$), such that
\beq
\mathcal V(x)^T C \mathcal V(x) = {\rm Id},
\eeq
where $V(x)^T$ denotes the transpose of the matrix $V(x)$.
Its coefficients are given by $C_{k,l;k',l'} = -\delta_{k,k'} X_{k,l+l'-1}$.
\el

\proof{From \eqref{xsumpoles} we have
\beq \label{EqC}
\frac{\x(z)-\x(z')}{z-z'}
= \sum_{k,l;k',l'} \frac{C_{k,l;k',l'}}{(z-\alpha_k)^{l}(z'-\alpha_{k'})^{l'}}
\quad , \quad
C_{k,l;k',l'} = -\delta_{k,k'} X_{k,l+l'-1}
\eeq
The matrix $C$ is made of triangular blocks because $X_{k,l+l'-1}=0$ if  $l+l'> d_k+1$. $C$ is invertible because the antidiagonals of each triangular block is $-C_{k,d_k}\neq 0$ by definition of $d_k$. We have :
\beq\label{eqxC1}
\forall\, \, 1\leq i,j\leq d\,\, :\,\, \left(\mathcal V(x)^T C \mathcal V(x')\right)_{i,j} = \frac{x-x'}{z^i(x)-z^j(x')}\,\frac{1}{\sqrt{\x'(z^i(x)) \x'(z^j(x'))}}
\eeq
Evaluating at $x=x'$ we get :
\beq
\mathcal V(x)^T C \mathcal V(x) = {\rm Id}.
\eeq
}

For example in the case $k=3$ and $(d_1,d_2,d_3)=(3,2,4)$, the matrix $C$ looks like :
\beq
C
=\begin{array}{|rr|rrrr|rrr|}
\hline
*&*&.&.&.&.&. &.&.\\
*&.&.&.&.&.&. &.&.\\
\hline
.&.&*&*&*& * &. &.&.\\
.&.&*&*& * &.&. &.&.\\
.&.&*& * &.&.&. &.&.\\
.&.& * &.&.&.&.&. &.\\
\hline
.&.&.&.&.&.&* &*&*\\
.&.&.&.&.&.&* &*&.\\
.&.&.&.&.&.&* &.&.\\
\hline
\end{array}
\eeq

Lemma \ref{lemmaVtCV} implies that :
\beq \label{Obs}
\mathcal V(x)^{-1} = \mathcal V(x)^T C \,\,\,\, \text{ and } \,\,\,\, 
\mathcal V(x) \mathcal V(x)^T = C^{-1}
\eeq
In particular, the matrix $C$ is always symmetric, and in each block it has the Hankel property : it depends only on $l+l'$.

\bc\label{corE}
The matrix $\mathcal V(x)^{-1} d\mathcal V(x)$ is antisymmetric, it is worth zero on its diagonal, and off diagonal entries are given by : 
\beq
\forall \,i\neq j\, :\, \left(\mathcal V(x)^{-1} d\mathcal V(x)\right)_{i,j}
= \frac{-\sqrt{dz^i(x)dz^j(x)}}{z^i(x)-z^j(x)}
= \frac{-1}{\mathcal E(z^i(x),z^j(x))}
\eeq
where $\mathcal E(z,z')=\frac{z-z'}{\sqrt{dz dz'}}$ is the prime form on the Riemann sphere.
\ec

\proof{
Taking the $x$-differential of \eqref{Obs} and using the fact that $C$ is independent of $x$ directly shows that $\mathcal V(x)^{-1} d\mathcal V(x)$ is antisymmetric.
Starting from \eqref{eqxC1} and differentiating with respect to $x'$ we get :
\bea
&&\left(\mathcal V(x)^{-1} d\mathcal V(x')\right)_{i,j} =  \frac{(x-x')\, dz^j(x')}{(z^i(x)-z^j(x'))^2}\,\frac{1}{\sqrt{\x'(z^i(x)) \x'(z^j(x'))}} \quad \quad\cr
&& - \frac{\x'(z^j(x'))\, dz^j(x')}{z^i(x)-z^j(x')}\,\frac{1}{\sqrt{\x'(z^i(x)) \x'(z^j(x'))}} \cr
&& -\frac{1}{2}\,\frac{\x''(z^j(x')) \, dz^j(x')}{\x'(z^j(x'))} \frac{x-x'}{z^i(x)-z^j(x')}\,\frac{1}{\sqrt{\x'(z^i(x)) \x'(z^j(x'))}}
\eea
We now take the limit $x\to x'$. Since $x=\x(z^i(x))=\x(z^j(x))$ we get the equalities $dx=\x'(z^i(x)) dz^i(x) = \x'(z^j(x)) dz^j(x)$. When $i\neq j$, the denominator does not vanish and only the terms without $x-x'$ in the numerator survive thus giving the claimed result. When $i=j$, the first two terms are computed by Taylor expansion up to the second order, i.e. involve the second derivative of $\x$, which is exactly canceled by the last term.
}

\bc\label{corB} With $G=GL_d(\mathbb C)$, with Cartan subalgebra $\Lieh$ the set of diagonal matrices, and defining the canonical basis of $\Lieh$ : $e_i=\text{diag}(0,\dots,0,\overset{i}{1},0,\dots,0)$, we have the identity :
\beq\label{VtoB}
\frac{\Tr \mathcal V(x) e_i \mathcal V(x)^{-1} \mathcal V(x') e_j \mathcal V(x')^{-1}}{(x-x')^2} dx dx'
 =  \frac{dz^i(x) dz^j(x')}{(z^i(x)-z^j(x'))^2} =  B(z^i(x),z^j(x')) ,
\eeq
where $B(z,z')=\frac{dz dz'}{(z-z')^2}$ is the fundamental $2^{\text{nd}}$ kind bi-differential of the Riemann sphere.
\ec

Moreover, we get the following property :
\begin{proposition}\label{PropV} The matrix $x\mapsto \mathcal V(x) e_i \mathcal V(x)^{-1}$ is a rational function of $z^i(x)$. 
It is only singular when $z^i(x)$ is at the branchpoints (i.e. finite branchpoints where $\x'(z)=0$ and poles of $\x(z)$ of degree at least $2$).
\end{proposition}

\proof{
Use $\mathcal V(x)^{-1} = \mathcal V(x)^T C$ and the definition of $\mathcal{V}(x)$ :
\beqq 
\left(\mathcal V(x) e_i \mathcal V(x)^T\right)_{(k,l),(k',l')}=\frac{1}{(z^i(x)-\alpha_k)^l(z^i(x)-\alpha_{k'})^{l'}\x'(z^i(x))}
\eeqq
This function has poles when $\x'(z^i(x))$ vanishes, i.e. at branchpoints, and also possibly at the punctures $z^i(x)=\alpha_k$.

If $\alpha_j$ is a puncture (i.e. a pole of $\x(z)$) but not a branchpoint we must have $d_j=1$ and thus $l=1$. We get that :
\beq
\left(\mathcal V(x) e_i \mathcal V(x)^T\right)_{(k,l),(k',l')}=O\left((z^i(x)-\alpha_j)^{d_j+1-l\delta_{k,j} -l'\delta_{k',j}}\right).
\eeq
The worst case happens when $k=k'=j$ implying $l=l'=1$, in which case the exponent is $0$, showing that $\mathcal V(x) e_i \mathcal V(x)^T$ has no pole.
}

We will now use the matrix $\mathcal{V}(x)$ and its properties to formulate our next assumption.

\subsection{Decomposition of the matrix of eigenvectors at order $\hbar^0$}

\begin{assumption}[Eigenvector decomposition]\label{AssumptionV}
We assume that there exists an invertible $d\times d$ matrix $v(t)$, independent of $x$, such that
\beq
V(x,t) = v(t) \mathcal V(x)
\eeq
is an invertible matrix whose columns are the eigenvectors of $L^{(0)}$ (and thus of $R^{(0)}$).
Consequently we have (not writing the $t$ dependence to lighten notations) :
\beq
L^{(0)}(x) = v \mathcal V(x) Y(x) \mathcal V(x)^T C v^{-1},
\eeq
\beq
R^{(0)}(x) = v \mathcal V(x) S(x) \mathcal V(x)^T C v^{-1} .
\eeq
In coordinates it is equivalent to :
\bea \label{Coord}
(L^{(0)}(x))_{i,j} &=&  \sum_{k,l,k',l',l'',m}  \frac{-\y(z^m(x)) v_{i;k,l} X_{k',l'+l''-1} (v^{-1})_{k',l'';j}}{(z^m(x)-\alpha_k)^l(z^m(x)-\alpha_{k'})^{l'} \x'(z^m(x))}\cr
(R^{(0)}(x))_{i,j} &=&  \sum_{k,l,k',l',l'',m}  \frac{-s(z^m(x)) v_{i;k,l} X_{k',l'+l''-1} (v^{-1})_{k',l'';j}}{(z^m(x)-\alpha_k)^l(z^m(x)-\alpha_{k'})^{l'} \x'(z^m(x))}
\eea

\end{assumption}

Notice that the last assumption implies that :
\bea
v(t)^{-1} L^{(0)}(x,t) v(t) C(t)^{-1} &\text{  and} \qquad v(t)^{-1} R^{(0)}(x,t) v(t) C(t)^{-1}\cr
\eea
are symmetric matrices.

\br
This is a very strong assumption on $L^{(0)}(x,t)$ and $R^{(0)}(x,t)$. It implies that the $x$-dependent part of $L^{(0)}(x,t)$ (resp. $R^{(0)}(x,t)$) has in fact only $\frac{d(d+1)}{2}$ degrees of freedom, rather than $d^2$. In other words it imposes $\frac{d(d-1)}{2}$ constraints on $L^{(0)}(x,t)$ (resp. $R^{(0)}(x,t)$).
\er

\begin{remark} The purpose of assumption \ref{AssumptionV} is to match the (defined below) correlator $W_2^{(0)}$ with the fundamental $2^{\text{nd}}$ kind bi-differential $B(z_1,z_2)$, defined in Corollary \ref{corB}, as it is necessary for the system to satisfy the topological type property.
\end{remark}

This assumption may look too restrictive on the matrices $L^{(0)}$ and $R^{(0)}$ but the set of matrices which satisfy it is far from empty. In fact most (if not all) well-known integrable systems satisfy it and examples of Painlev\'{e} systems and $(p,q)$ minimal models are given in appendix \ref{AppendixExamples}.

\subsection{Classification of admissible systems}

From \eqref{Coord} we must have :
\bea\label{decompL0residues}
\left(v^{-1} L^{(0)}(x) v C^{-1}\right)_{k,l;k',l'} 
&=& \sum_{j=1}^d \mathcal V_{k,l}(z^j(x)) \mathcal V_{k',l'}(z^j(x)) \,\y(z^j(x)) \cr
&=& \sum_{j=1}^d \frac{1}{(z^j(x)-\alpha_k)^l}\,\frac{1}{(z^j(x)-\alpha_{k'})^{l'}}  \,\frac{\y(z^j(x))}{\x'(z^j(x))} \cr
&=& \sum_{j=1}^d \Res_{z\to z^j(x)} \frac{1}{(z-\alpha_k)^l}\,\frac{1}{(z-\alpha_{k'})^{l'}}  \,\frac{\y(z)}{\x(z)-x} \cr
&=& -\sum_{p\in \{ {\rm poles\,of\,}\,\x\,{\rm and}\,\y\}} \Res_{z\to p} \frac{1}{(z-\alpha_k)^l}\,\frac{1}{(z-\alpha_{k'})^{l'}}  \,\frac{\y(z)}{\x(z)-x} \cr
\eea

The pole at $z=\alpha_i$ gives a polynomial of $x$ of degree lower or equal to  $\frac{l\delta_{k,i}+l'\delta_{k',i}-2d_i+\deg_{\alpha_i}\y}{d_i}$. Thus if $\y$ has no pole at $\alpha_i$, this gives at most an $x$ independent term, and only for $k=k'=i$, $l=l'=d_i$.

If $p$ is a pole of $\y$ which is not a pole of $\x$, we get a pole $(\x(p)-x)^m$ with $m\leq  \frac{\deg_p \y}{1+{\rm ord}_p \x'}$. 

\subsubsection{Decomposition on  $z^r$}

Any rational function $\y(z)$ can be uniquely written as
\beq
\y(z) = \sum_{r=0}^{d-1} z^r f_r(\x(z)).
\eeq
where $f_r(x)$ is a rational function of $x$. Since functions of $x$ go through \eqref{decompL0residues}, it is sufficient to study the cases $\y(z)=z^r$.

So let us substitute $\y(z)\to z^r$ in \eqref{decompL0residues}, with $0\leq r\leq d-1$, and
we assume (up to a M\"obius transformation of $z$) that $\x$ is regular at $z=\infty$ (i.e. none of the $\alpha_i$'s is located at $\infty$).
The contribution to \eqref{decompL0residues} of poles at $\alpha_i$'s is a constant matrix $\hat A_{i,r}$, which is a triangular block of size $d_i$, which we denote :
\beq
A_{r,0} = \sum_i \tilde A_{i,r}
\qquad , \qquad 
(\tilde A_{i,r})_{k,l;k',l'}= \delta_{k,i}\delta_{k',i}  \tilde A_{i,r,l+l'}
\eeq
that is non vanishing only if $l+l'\geq d_i+1$. On the anti-diagonal we get :
\beq
\tilde A_{i,r,d_i+1} = \frac{-\alpha_i^r}{X_{i,d_i}}.
\eeq
Example :
\beq
\tilde A_{2,r}
=\begin{array}{|rr|rrrr|rrr|}
\hline
.&.&.&.&.&.&. &.&.\\
.&.&.&.&.&.&. &.&.\\
\hline
.&.&.&.&.& * &. &.&.\\
.&.&.&.& * &*&. &.&.\\
.&.&.& * &*&*&. &.&.\\
.&.& * &*&*&*&.&. &.\\
\hline
.&.&.&.&.&.&. &.&.\\
.&.&.&.&.&.&. &.&.\\
.&.&.&.&.&.&. &.&.\\
\hline
\end{array}
\eeq
The contribution of the pole at $z=\infty$ takes the form :
\beq
\sum_{m=1}^{r} \frac{\hat A_{r,m}}{(x-\x(\infty))^{m}}
\eeq
and we have that :
\beq
(\hat A_{r,m})_{k,l;k',l'} = 0 \quad {\rm if}\quad l+l'-2>r-m.
\eeq
For example if $r=1$ we have $(\hat A_{1,1})_{k,l'k',l'} = \delta_{l,1}\delta_{l',1}$ :
\beq
\hat A_{1,1}
=\begin{array}{|rr|rrrr|rrr|}
\hline
1&.&1&.&.&.&1 &.&.\\
.&.&.&.&.&.&. &.&.\\
\hline
1&.&1&.&.&.&1 &.&.\\
.&.&.&.& .&.&. &.&.\\
.&.&.& . &.&.&. &.&.\\
.&.& . &.&.&.&.&. &.\\
\hline
1&.&1&.&.&.&1 &.&.\\
.&.&.&.&.&.&. &.&.\\
.&.&.&.&.&.&. &.&.\\
\hline
\end{array}
\eeq

Finally :
\beq
v^{-1} L^{(0)}(x) v C^{-1}
=   \sum_{r=0}^{d-1} \sum_{m=0}^{r} \frac{f_{r}(x)}{(x-\x(\infty))^m}\, \hat A_{r,m},
\eeq
we end up with a matrix $L^{(0)}(x)$ that, up to some left/right multiplications by $x$-independent matrices ($v$ on the left and $Cv^{-1}$ on the right) of a very restrictive form. 

\subsubsection{Decomposition on $(z-\alpha_i)^{-r}$}

A better decomposition is the following :
any function $\y(z)$ can be uniquely written as
\beq
\y(z) = 
\sum_{i} \sum_{r=1}^{d_i} \frac{\mathcal Y_{i,r}(\x(z))}{(z-\alpha_i)^r},
\eeq
where each $\mathcal Y_{i,r}(x)$ is a rational function of $x$, given by
\beq
\mathcal Y_{i,r}(x) = - \sum_j \frac{\y(z^j(x))}{\x'(z^j(x))}\,\sum_{l=r}^{d_i} X_{i,l} (z^j(x)-\alpha_i)^{r-l-1}
\eeq

This gives
\beq
v^{-1} L^{(0)}(x) v C^{-1}
= \sum_{i,r} \mathcal Y_{i,r}(x) A_{i,r}(x)
\eeq
where the matrices $A_{i,r}(x)$ are computed using  $\y(z)=(z-\alpha_i)^{-r}$ with $1\leq r\leq d_i$.
Using \eqref{decompL0residues}, we find that each $A_{i,r}(x)$ is a polynomial of $x$ of degree at most $1$
\beq
A_{i,r}(x) =  x A'_{i,r} + A_{i,r},
\eeq
where the matrices $A'_{i,r}$ and $A_{i,r}$ have the following block shape :
\beqq
r=1 \quad \to \quad
A_{2,1}
=\begin{array}{|rr|rrrr|rrr|}
\hline
.&.&.&.&.&*&. &.&.\\
.&.&.&.&.&*&. &.&.\\
\hline
.&.&.&.&*&*&. &.&.\\
.&.&.&*&*&*&. &.&.\\
.&.&*&*&*&*&. &.&.\\
*&*&*&*&*&*&*&* &*\\
\hline
.&.&.&.&.&*&. &.&.\\
.&.&.&.&.&*&. &.&.\\
.&.&.&.&.&*&. &.&.\\
\hline
\end{array}
\quad , \quad
A'_{2,1}
=\begin{array}{|rr|rrrr|rrr|}
\hline
.&.&.&.&.&.&. &.&.\\
.&.&.&.&.&.&. &.&.\\
\hline
.&.&.&.&.&.&. &.&.\\
.&.&.&.&.&.&. &.&.\\
.&.&.&.&.&.&. &.&.\\
.&.&.&.&.&*&.&. &.\\
\hline
.&.&.&.&.&.&. &.&.\\
.&.&.&.&.&.&. &.&.\\
.&.&.&.&.&.&. &.&.\\
\hline
\end{array}
\eeqq
\beqq
r=2 \quad \to \quad
A_{2,2}
=\begin{array}{|rr|rrrr|rrr|}
\hline
.&.&.&.&*&*&. &.&.\\
.&.&.&.&*&*&. &.&.\\
\hline
.&.&.&*&*&*&. &.&.\\
.&.&*&*&*&*&. &.&.\\
*&*&*&*&*&*&* &*&*\\
*&*&*&*&*&*&*&* &*\\
\hline
.&.&.&.&*&*&. &.&.\\
.&.&.&.&*&*&. &.&.\\
.&.&.&.&*&*&. &.&.\\
\hline
\end{array}
\quad , \quad
A'_{2,2}
=\begin{array}{|rr|rrrr|rrr|}
\hline
.&.&.&.&.&.&. &.&.\\
.&.&.&.&.&.&. &.&.\\
\hline
.&.&.&.&.&.&. &.&.\\
.&.&.&.&.&.&. &.&.\\
.&.&.&.&.&*&. &.&.\\
.&.&.&.&*&*&.&. &.\\
\hline
.&.&.&.&.&.&. &.&.\\
.&.&.&.&.&.&. &.&.\\
.&.&.&.&.&.&. &.&.\\
\hline
\end{array}
\eeqq
and so on, for higher $r$, the non-vanishing off-diagonal blocks have size $r\times d_i$, and the non-vanishing entries are some universal functions of the $X_{i,k}$'s.
Eventually we have
\beq
v^{-1} L^{(0)}(x) v C^{-1}
= \sum_{i,r} \mathcal Y_{i,r}(x) (xA'_{i,r}+A_{i,r})
\eeq

Again we obtain a very restrictive class of matrices $L^{(0)}(x)$.

\subsubsection{Classification of $R^{(0)}(x)$}

The previous results also hold for $R^{(0)}$ with $\y$ replaced by $\s$. However due to the requirement that the auxiliary curve does not have any double points, the interesting cases are even more restrictive.

We may uniquely write
\beq
\s(z) = \sum_{j=0}^{m} f_j(\x(z)) \,z^j
\qquad , \qquad m\leq d-1.
\eeq
%
%
%
If $m=1$, then it is obvious that there can be no double points, in that case
\beq
\s(z)=f_0(\x(z)) + f_1(\x(z)) z.
\eeq
In other words, $R^{(0)}(x,t)$ 
\bea
R^{(0)}(x,t)
&=& f_0(x,t) v(t) \hat A_{0,0}(t) C(t) v(t)^{-1} \cr
&& + f_1(x,t) \,v(t)\left( \hat A_{1,0}(t)C(t) + \frac{\hat A_{1,1}(t) C(t)}{x-\x(\infty,t)}\right)v(t)^{-1}.
\eea
Up to a M\"obius transformation on $z$ we could have chosen $z=\infty$ to be a pole of $\x$, and then we would have obtained
\beq
R^{(0)}(x,t) 
= f_\infty(x,t)  \,\,v(t)\,(A_{\infty,1}(t) C(t)+x A'_{\infty,1}(t)C(t) ) v(t)^{-1}.
\eeq
Remark that all $(p,q)$ minimal models, as well as Painlev\'e systems are indeed of this form.

Notice that if $d_\infty>1$ then $A'_{\infty,1}(t)C(t)$ is a nilpotent matrix :
\beq
\left(A'_{\infty,1}(t)C(t) \right)_{k,l;k',l'} = \frac{1}{X_{\infty,d_\infty}}\,\delta_{k,\infty}\delta_{k',\infty} \delta_{l,d_\infty} \delta_{l',1}.
\eeq

%
%
%
%

%


\subsection{Assumptions regarding the $\hbar$ higher orders}

In order to prove the topological type property and in addition to assumptions \ref{genuszeroassumption} and \ref{AssumptionV}, we make the following sufficient assumptions regarding the spectral curve and the possible singularities of the system. We shall need the notion that $L^{(k\geq 1)}$ has to be ``less singular'' than $L^{(0)}$ -- symbolically denoted $L^{(k)} \prec L^{(0)}$ --. Our precise statement is the following:

\begin{assumption}[Analytic behavior $L^{(k)} \prec L^{(0)}$]\label{Asspolesordrek}
We assume that:
\begin{itemize}
\item for every $k\geq 1$, all poles of $L^{(k)}(x,t)$ are among the poles of $L^{(0)}(x,t)$.
\item for any matrix $\tilde C$, and any generic distinct $x_0,x_1$, the following $\hbar$-formal series whose coefficients are bi-rational functions of $x$ and $y$:
\beq
\frac{\det\left(y-L(x)-\frac{\tilde C}{(x-x_0)(x-x_1)}\right)
-\det\left(y-L^{(0)}(x)\right)}{E_y(x,y)}\,dx
\eeq
is, when restricted to the spectral curve, a one-form $\Omega(z)$ that is analytic (at each order in $\hbar$) at all singularities of $L^{(0)}(x)$.

Equivalently, its only singularities may either be poles over $x=x_0$ and $x=x_1$, due to the $\frac{\tilde C}{(x-x_0)(x-x_1)}$ term, or at double points of $\spcurve$: $(b_i,\tilde b_i)$.
\bea
\Omega(z)={\det\left(\y(z)-L(\x(z))-\frac{\tilde C}{(\x(z)-x_0)(\x(z)-x_1)}\right)}\,\frac{d\x(z)}{E_y(\x(z),\y(z))}  \cr
=
\sum_i \beta_i \left(\frac{dz}{z-b_i} - \frac{dz}{z-\tilde b_i}\right)
+ \sum_{i\in \{0,1\}} \sum_{j=1}^d\sum_{k=1}^d \frac{c_{i,j,k} dz}{(z-z^j(x_i))^k}
\eea
where the coefficients $\beta_i$, $c_{i,j,k}$ are formal power series of $\hbar$, starting at $O(\hbar)$.

\end{itemize}
\end{assumption}

In other words, the $\hbar$ correction terms do not change the Newton's polygon of $E(x,y)$. They may only change the interior coefficients, as well as possibly adding poles over $x=x_{0}$ or $x=x_1$. 

Evaluating this one-form at $z^i(x)$, inserting and subtracting the diagonal term $Y(x)=V(x)^{-1}L^{(0)}(x)V(x)$ and then expanding the determinant, we get after simplification that it is equal to
\beq
\Omega(z^i(x)) = \sum_{I\subset \{1,\dots,d\},\, i\in I} \frac{\underset{I\times I}{\det} \left( V(x)^{-1}\left(L(x)-L^{(0)}(x)+\frac{\tilde C}{(x-x_0)(x-x_1)}\right)V(x) \right)}{\underset{j\in I,\, j\neq i}{\prod} (\y(z^i(x))-\y(z^j(x)))} \,\, dx .
\eeq
In particular, to order $\hbar$ we must have $\forall\,i$ :
\beq
 \left( V(x)^{-1}L^{(1)}(x)V(x) \right)_{i,i} \,dx =0
\eeq
(which implies $W_1^{(0)}(x.e_i)=0$, as we will see below).
This is equivalent to say that $L^{(1)}(x)$ must be derived from $L^{(0)}(x)$, i.e. $\exists\, \tilde L^{(1)}(x)$ such that
\beq
L^{(1)}(x) = [\tilde L^{(1)}(x),L^{(0)}(x)].
\eeq
At order $\hbar^2$ we get that
\beqq
\left( V(x)^{-1}L^{(2)}(x)V(x) \right)_{i,i}\, dx   
- \sum_{j\neq i} \frac{ \left( V(x)^{-1}L^{(1)}(x)V(x) \right)_{i,j} \left( V(x)^{-1}L^{(1)}(x)V(x) \right)_{j,i}}{ (\y(z^i(x))-\y(z^j(x)))}\,dx  
\eeqq
is analytic at all poles of $\y$.

\br
Assumption \ref{Asspolesordrek} may appear technical but it can be proved easily in many cases. For example :
\begin{enumerate}

\item Assumption \ref{Asspolesordrek} is trivially verified if for all $k\geq 1$, $L^{(k)}$ is independent of $x$. (This happens for the Airy Lax pair for example).

\item Assumption \ref{Asspolesordrek} is verified if $L(x,\hbar)$ is a Fuchsian system, i.e. has only simple poles $c_i(t)$ independent of $\hbar$, and residues $C_i(t,\hbar)$ whose eigenvalues are independent of $\hbar$:
\beq 
L(x,t,\hbar)=\sum_{i=1}^p \frac{C_i(t,\hbar)}{x-c_i(t)} 
\eeq
Indeed, in that case the poles of $L^{(k)}$ are the same as those of $L^{(0)}$.
The eigenvalues of $L(x,t,\hbar)$ have only simple poles above $x=c_i(t)$, with residues the eigenvalues of $C_i(t,\hbar)$, and thus all the singular behavior of the eigenvalues of $L(x,t,\hbar)$, is independent of $\hbar$, showing that the characteristic polynomials of $L(x,t,\hbar)$ and $L^{(0)}(x,t)$ may only differ from the interior of their Newton's polygon.
\end{enumerate} 

\er

\subsection{Parity Assumption}

In order to prove sufficient conditions for the topological type property, we need (as proposed in \cite{BBE14}) another assumption :

\begin{assumption}[Parity]\label{Asssym}
We assume that there exists a matrix $\Gamma(t,\hbar)=\underset{k=0}{\overset{\infty}{\sum}} \hbar^k \Gamma^{(k)}(t)$, independent of $x$, such that
\beq \label{Parity1}
L(x,t,-\hbar) = \Gamma(t,\hbar)^{-1} L(x,t,\hbar)^T \Gamma(t,\hbar).
\eeq
with
\beq \label{Gamma0}
\Gamma^{(0)} = (v^T(t))^{-1} C v(t)^{-1} = \Gamma^{(0)\,T}.
\eeq
\end{assumption}

Again this assumption is not empty and it is satisfied for many well-known integrable systems. For example it is satisfied for the Painlev\'{e} Lax pairs and the $(p,q)$ minimal models. 
Also,  to leading order in $\hbar$ this assumption is a consequence of assumption \ref{AssumptionV}.

This assumption was made in \cite{BBE14} and automatically gives the parity condition of the TT property. This assumption is not known to be necessary, but so far we have not found any counter example in the literature.

\medskip 
Notice that we have
\beq
\Gamma(t,-\hbar)  = \Gamma(t,\hbar)^T,
\eeq
i.e. for all $k\geq 0$ :
\beq
\Gamma^{(k)}(t) = (-1)^k\,\Gamma^{(k)}(t)^T.
\eeq
In other words, the coefficients of the matrices appearing in the series expansion of $\Gamma(t,\hbar)$ are either symmetric or antisymmetric matrices depending on the parity of their index.  

\section{The matrix $M(x.E,t)$ and the correlators $W_n$ \label{SectionM}}

Following the works of \cite{BE09,BBE14} we now define the following quantities (we omit the $t$-dependence for clarity) :

\bd\label{DefM}
For any solution $\Psi(x)$ of the system \eqref{GeneralLax}, and any constant $d\times d$ matrix $E\in \Lieg$, we define
\beq
M(x.E) = \Psi(x) E \Psi(x)^{-1}.
\eeq
It satisfies the adjoint system to \eqref{GeneralLax}
\bea\label{LaxM}
\hbar \partial_x M(x.E) &=& [L(x),M(x.E)] \cr
\hbar \partial_t M(x.E) &=& [R(x),M(x.E)] .
\eea
In other words, at fixed $E$, the map $x\mapsto M(x.E)$ is a flat section of the adjoint connection on the adjoint bundle.
\ed

\br
Equations \eqref{LaxM} are isospectral, i.e. they imply that the eigenvalues of $M(x.E)$ are independent of $x$ and $t$.

\er

Most often we shall choose $E$ in a Cartan subalgebra $\Lieh\subset \mathfrak{gl}_d(\mathbb C)$, i.e. a diagonal matrix, and thus define
 $e_a$ the basis of rank one diagonal projectors :
\beq
e_a=\text{diag}(0,\dots,0,\overset{a}{1},0,\dots,0).
\eeq
In that case, since $e_a$ is a rank one projector, then so is $M(x.e_a)$ :
\beq
M(x.e_a,t,\hbar)^2=M(x.e_a,t,\hbar)
\qquad , \qquad
\operatorname{Tr} M(x.e_a,t,\hbar)=1.
\eeq

\subsection{WKB expansion}

WKB expansions are usually defined for wave functions $\Psi(x)$, but here we shall use the adjoint version of WKB for $M(x.E)$. In the end, the two versions are equivalent, as explained in remark \ref{rmkequWKBPsiM} below.

Note that WKB expansions are defined only within sectors, where $x$ belongs to an open simply connected domain containing no singularity of $\x,\y,\s$ neither any branchpoints. In such sectors, a consistent analytic ordering of preimages $z^1(x),\dots,z^d(x)$ is well defined, as well as the sign of the square-root $\sqrt{\x'(z^i(x))}$.

The system \eqref{LaxM} admits an $\hbar$ formal series solution:

\begin{theorem}[$\hbar$ expansion of $M$]\label{Mexp}  
There exists a unique $\hbar$-formal series expansion 
\bea \label{Mexpp} M(x.e_a,t,\hbar)
= V(x,t) e_a V(x,t)^{-1} + \sum_{k=1}^\infty M^{(k)}(z^a(x),t)\hbar^k
\eea
that satisfies the differential system :
\bea \label{SystemM}
\hbar \partial_x M(x.e_a,t,\hbar)&=&\left[L(x,t,\hbar),M(x.e_a,t,\hbar)\right] \cr
 \hbar \partial_t M(x.a_a,t,\hbar)&=&\left[R(x,t,\hbar),M(x.e_a,t,\hbar)\right]
 \eea
and such that $M(x.e_a,t,\hbar)$ is a rank one projector :
\beq\label{Mproj}
M(x.e_a,t,\hbar)^2=M(x.e_a,t,\hbar)
\qquad , \qquad
\operatorname{Tr} M(x.e_a,t,\hbar)=1.
\eeq
Moreover, the coefficients $(M^{(k)}(z,t))_{i,j}$ are \textbf{rational functions of $z$}.
\end{theorem}

\br\label{rmkequWKBPsiM} The $\hbar$-expansion of $M$ is equivalent to the WKB expansion for $\Psi$ given by :
\beq\label{WKBpsi}
\Psi^{\rm WKB}(x,t,\hbar) = V(x,t) \left({\mathbf 1}_d + \sum_{k\geq 1} \hbar^k \Psi^{(k)}(x,t) \right) e^{\hbar^{-1} T(x,t)} 
\eeq
where $T(x,t) = {\rm diag}(T_1(x,t),\dots,T_d(x,t))$ with
\beq
\partial_x T_i(x,t) = y_i(x,t)
\quad , \quad
\partial_t T_i(x,t) = s_i(x,t).
\eeq
Indeed, if one chooses $E=e_a$ diagonal, the exponential terms disappear in the product $M=\Psi E \Psi^{-1}$ and one finds an $\hbar$ expansion for $M$ without exponential terms.

Vice-versa, $\Psi$ is recovered from $M$ by the formula (proved in appendix \ref{appequWKBPsiM})
\beq
\Psi(x,t,\hbar)_{i,a} = M_{i,1}(x.e_a)\, e^{\frac{1}{\hbar} \int^x \frac{\underset{k=1}{\overset{d}{\sum}} M_{1,k}(x'.e_a) L_{k,1}(x')}{M_{1,1}(x'.e_a)}}dx'
\eeq
and an $\hbar$-expansion for $M$ of the form \eqref{Mexpp} leads to a WKB type expansion for $\Psi$.
\er

\proof{First notice that the property of being a rank one projector is compatible with the flows (in $x$ and $t$) of the differential system. Indeed, the flows are isospectral, meaning that the eigenvalues of $M(x.E)$ are preserved.

Let us start by studying the formal expansion of $M$ conjugated by $V$, i.e. $\tilde M(x.e_a,t,\hbar)  = V(x,t)^{-1} M(x.e_a,t,\hbar) V(x,t)$. Its $\hbar$-expansion is of the form :
\beq
\tilde M(x.e_a,t,\hbar)= e_a  + \sum_{k=1}^\infty \hbar^k \, \tilde M^{(k)}(x.e_a,t).
\eeq
It satisfies the differential system :
\bea \label{SystemtdM} 
\hbar \partial_t \td{M}(x.e_a,t)
&=&\left[V^{-1}(x,t)R(x,t)V(x,t)-V(x,t)^{-1}(\hbar \partial_t V(x,t)),\td{M}(x.e_a,t)\right]\cr
\hbar \partial_x \td{M}(x.e_a,t)
&=&\left[V^{-1}(x,t)L(x,t)V(x,t)-V(x,t)^{-1}(\hbar \partial_x V(x,t)),\td{M}(x.e_a,t)\right]\cr
\eea
The differential system \eqref{SystemtdM} for $\td M$ is sufficient to determine recursively the coefficients $\td{M}^{(k)}(x.e_a,t)$ of the expansion. 
Let us first denote
\beq
\tilde U(x) = V(x,t)^{-1} \partial_t V(x,t) =  \mathcal V^T C v^{-1}\partial_t v \mathcal V + \mathcal V^{-1} \partial_t\mathcal V,
\eeq
and
\beq
\tilde U(x)_{i,j} = \tilde u(z^i(x),z^j(x)), i\neq j
\quad , \quad
\tilde U(x)_{i,i} = \tilde u_{\rm diag}(z^i(x)) 
\eeq
where $\tilde u(z,z')\sqrt{\x'(z)\x'(z')}$ is a rational function of $z$ and $z'$, and $\tilde u_{\rm diag}(z)\x'(z)$ is a rational function of $z$.
Similarly, we define :
\beq
U(x) = V(x,t)^{-1} \partial_x V(x,t)) =   \mathcal V^T C \partial_x\mathcal V,
\qquad U(x)_{i,j} = u(z^i(x),z^j(x)),\,i\neq j
\eeq
According to corollary \ref{corE}, we have $U(x)_{i,i}=0$ and if $i\neq j$
\beq
U(x)_{i,j} = \frac{-1}{(z^i(x)-z^j(x))\,\sqrt{\x'(z^i(x))\x'(z^j(x))}}.
\eeq

\medskip

The first step of the proof is to show by recursion on $k$ that :
\begin{itemize}
\item if $i\neq j$,  $\td{M}^{(k)}(x.e_a,t)_{i,j} = m_k(z^a(x),z^i(x),z^j(x))_{i,j}$, where $m_{k}(z,z',z'')_{i,j} \sqrt{\x'(z')\x'(z'')}$ is a rational function of all its arguments. 
\item if $i=j$,  $\td{M}^{(k)}(x.e_a,t)_{i,i} = m_k(z^a(x),z^i(x))_{i,i}$ where $m_{k}(z,z')_{i,i} \x'(z')$ is a rational function of all its arguments. For convenience in the notations, we shall write $m_k(z,z')_{i,j} = m_k(z,z',z')_{i,i}$.
\end{itemize}

To leading order in $\hbar$, equations \eqref{SystemtdM} reduce to :
\beq
[S,\tilde M^{(0)}] =0= [Y,\tilde M^{(0)}]
\eeq
Thus, $\tilde M^{(0)}(x.e_a)=e_a$ satisfies the last equations and moreover we have $m_{0}(z,z',z'')_{i,j} = 0$ if $i\neq j$ and $m_{0}(z,z',z')_{i,i} = \delta_{i,a}$ if $i=j$. Consequently the induction is initialized for $k=0$.

Let us look at $k\geq 1$. Looking at the first equation of \eqref{SystemtdM} at order $\hbar^k$ provides:
\beq \partial_t \td{M}^{(k-1)}=\left[S,\td{M}^{(k)}\right]+\sum_{l=1}^k \left[V^{-1}R^{(l)}V,\td{M}^{(k-l)}\right]+\left[\td{M}^{(k-1)},V^{-1}\partial_t V\right]\eeq
In other words for $i\neq j$ we get :
\bea \label{OffDiagonalPart}
&& \left(\td{M}^{(k)}(x.e_a,t)\right)_{i,j} \cr
&=& \frac{1}{s_i(x)-s_j(x)}\Big( \partial_t \left(\td{M}^{(k-1)}\right)_{i,j}\cr
&&-\sum_{l=1}^k\left[V^{-1}R^{(l)}V,\td{M}^{(k-l)}\right]_{i,j}+\left[\td{M}^{(k-1)},V^{-1}\partial_t V\right]_{i,j}\Big) \cr
&=& \frac{1}{s(z^i(x))-s(z^j(x))}\Big( \partial_t m_{k-1}(z^a(x),z^i(x),z^j(x))_{i,j}\cr
&&-\sum_{l=1}^k\left[\mathcal V^T v^T C R^{(l)} v \mathcal V,\td{M}^{(k-l)}\right]_{i,j}+\left[\td{M}^{(k-1)},\tilde U(x) \right]_{i,j}\Big) \cr
&=& \frac{1}{s(z^i(x))-s(z^j(x))}\Big( \partial_t m_{k-1}(z^a(x),z^i(x),z^j(x))_{i,j}\cr
&&-\sum_{l=1}^k \sum_{p,q,r} \mathcal V_p(z^i(x)) (v^T C R^{(l)}(x) v)_{p,q} \mathcal V_q(z^r(x)) m_{k-l}(z^a(x),z^r(x),z^j(x))_{r,j} \cr
&&+\sum_{l=1}^k \sum_{p,q,r}  m_{k-l}(z^a(x),z^i(x),z^r(x))_{i,r}\mathcal V_p(z^r(x)) (v^T C R^{(l)}(x) v)_{p,q} \mathcal V_q(z^j(x))  \cr
&&-\sum_r m_{k-1}(z^a(x),z^i(x),z^r(x))_{i,r} \tilde u(z^r(x),z^j(x))   \cr
&&+\sum_r \tilde u(z^i(x),z^r(x))  m_{k-1}(z^a(x),z^r(x),z^j(x))_{r,j}  \Big) 
\eea
To see that, upon multiplying by $\sqrt{\x'(z^i(x))\x'(z^j(x))}$ it is a rational function of $z^a(x),z^i(x),z^j(x)$, notice that :

- The square roots contained in $\mathcal V_q(z^r)$ get multiplied by square roots from the $m_{k-l}(z^r)$, and so all square roots come by pairs, thus providing rational functions of $z$.

- We can replace $x$ in $R^{(l)}(x)$ by $R^{(l)}(\x(z^a(x))$ which is a rational function of $z^a(x)$.

- For any rational function $f(z)$, the sum $\underset{r}{\sum} f(z^r(x))$ is a a symmetric function of $z^1(x),\dots,z^d(x)$, therefore it is a rational function of their elementary symmetric polynomials, i.e. of the coefficients of $\x(z)-x$. Consequently it is a rational function of $x$, hence it is a rational function of $\x(z^a(x))$. Thus, we end up with a rational function of $z^a(x)$.

For the diagonal entries $i=j$, we use the fact that $M^2=M$ implying that $\td{M}^2=\td{M}$ too. Consequently, we have $\td{M}_{i,i}=(\td{M})_{i,i}^2+\underset{j\neq i}{\sum}\td{M}_{i,j}\td{M}_{j,i}$ and thus looking at order $\hbar^k$ we get:
\beq \left(1-2(\td{M}^{(0)})_{i,i}\right)\left(\td{M}^{(k)}\right)_{i,i}=\sum_{l=0}^k\sum_{i\neq j} (\td{M}^{(l)})_{i,j}(\td{M}^{(k-l)})_{j,i}-\sum_{l=1}^{k-1}(\td{M}^{(l)})_{i,i}(\td{M}^{(k-l)})_{i,i}
\eeq
By definition we have $(\td{M}^{(0)})_{i,i}=\delta_{i,a}\in\{0,1\}$, thus we have $1-2(\td{M}^{(0)})_{i,i}=\pm 1$ and:
\bea \label{DiagonalPart} 
\left(\td{M}^{(k)}(x.e_a,t)\right)_{i,i}
&=&\frac{1}{1-2\delta_{i,a}}\Big( \sum_{l=0}^k\sum_{i\neq j} (\td{M}^{(l)}(x.e_a,t))_{i,j}(\td{M}^{(k-l)}(x.e_a,t))_{j,i}\cr
&&-\sum_{l=1}^{k-1}(\td{M}^{(l)}(x.e_a,t))_{i,i}(\td{M}^{(k-l)}(x.e_a,t))_{i,i}\Big) \cr
&=&\frac{1}{1-2\delta_{i,a}}\Big( \sum_{l=0}^k\sum_{i\neq j} \sum_{r=1}^d 
m_l(z^a,z^i,z^r)_{i,r} m_{k-l}(z^a,z^r,z^j)_{r,j} ) \cr
&&-\sum_{l=1}^{k-1} \sum_{r=1}^d m_l(z^a,z^i,z^i)_{i,i} m_{k-l}(z^a,z^i,z^i)_{i,i}\Big) 
\eea
which is also a rational function of $z^a,z^i$. This proves the recursion for $\tilde M$. We shall now use this result to prove that the coefficients of $M^{(k)}$ are rational functions of $z$. Conjugating the last result by $V(x,t)$ gives:
\beq
M^{(k)}_{i,j} = \sum_{p,q,n,r=1}^d v_{i,n} \mathcal V_n (z^p(x)) m_k(z^a,z^p,z^q)_{p,q} \mathcal V_r(z^q) (Cv)_{r,j}
\eeq
The sum over $p$ and $q$ yields a rational function of $z^a(x)$ and thus the coefficients of $M^{(k)}$ are rational functions of $z$. 
} 

\medskip

Note that similar computations with the $x$-differential equation lead to :
\bea \label{AlternativeL}
\left(\td{M}^{(k)}(x.E,t)\right)_{i,j}&=& \frac{1}{y_i(x)-y_j(x)}\Big( \partial_x \left(\td{M}^{(k-1)}\right)_{i,j}\cr
&&-\sum_{l=1}^k\left[V^{-1}L^{(l)}V,\td{M}^{(k-l)}\right]_{i,j}+\left[\td{M}^{(k-1)},V^{-1}\partial_x V\right]_{i,j}\Big)\cr
\left(\td{M}^{(k)}(x.E,t)\right)_{i,i}&=&\frac{1}{1-2\delta_{i,a}}\Big( \sum_{l=0}^k\sum_{j\neq i} (\td{M}^{(l)}(x.E,t))_{i,j}(\td{M}^{(k-l)}(x.E,t))_{j,i}\cr
&&-\sum_{l=1}^{k-1}(\td{M}^{(l)}(x.E,t))_{i,i}(\td{M}^{(k-l)}(x.E,t))_{i,i}\Big)
\eea

We shall use these results to analyze the possible singularities of the matrices $M^{(k)}(z,t)$. The results are presented in the following section.

\subsection{Singularity structure of $M$}

We have the following theorem :
\begin{theorem}[Singularity structure of $M$]\label{TheoM} 
The matrices $\left(M^{(k)}(z)\right)_{k\geq 0}$ may only have poles at the branchpoints or at the poles of $L^{(0)}(x)$. In particular, they are regular at the double points of $\spcurve$.
\end{theorem}

\proof{
Let us prove the theorem by recursion on $k$. For $k=0$ we have :
\beq
\left(v^{-1} M^{(0)}(z) v C^{-1}\right)_{i,l;j,l'} =  \frac{1}{(z-\alpha_i)^l\,(z-\alpha_{j})^{l'}\,\x'(z)}
\eeq
Thus, $M^{(0)}(z)$ may only have poles at the zeros of $\x'(z)$ (that are branchpoints) or at the $\alpha_i$'s. In particular $\left(v^{-1} M^{(0)}(z) v C^{-1}\right)_{i,l;j,l'}$ has a pole at $z=\alpha_m$ of degree equal to $l\delta_{m,i} + l' \delta_{m,j} - d_m-1$. The only case for which $\alpha_m$ is not a branchpoint corresponds to $d_m=1$. In that case, we necessarily have $l=l'=1$ and the degree of the pole is zero. In other words if $\alpha_m$ is not a branchpoint, $M^{(0)}(z)$ has no pole at $z=\alpha_m$. Therefore $M^{(0)}(z)$ has poles only at branchpoints.

Then, assume that $M^{(k')}(z)$ has poles only at branchpoints and/or poles of $L^{(0)}(x)$ for all $k'<k$.
By contradiction, let us assume that $M^{(k)}(z)$ has a pole at a point $p$ of some order $r\geq 1$ where $p$ is not a branchpoint nor a pole of $L^{(0)}$. We write
\beq
M^{(k)}(z) \overset{z\to p}{=} \frac{M^{(k),r}}{(z-p)^r} + O\left((z-p)^{1-r}\right)
\eeq
The polar part at $z=p$ of the equation $M=M^2$ at order $\hbar^k$, is :
\beq \label{Polpart}
M^{(k),r} = M^{(k),r}(z) M^{(0)}(p) + M^{(0)}(p) M^{(k),r}.
\eeq
Notice that $M^{(0)}(p) $ is a rank one matrix of the form
\beq
v^{-1} M^{(0)}(p) v = u u^T C 
\qquad , \quad u^T C u=e_a \text{  with  } u=\mathcal{V}(x)e_a=v^{-1}Ve_a
\eeq
Moreover, $v u=Ve_a$ (resp. $e_a u^T C v^{-1}$) is a right (resp. left) eigenvalue of $R^{(0)}(\x(p))$ of eigenvalue $\s(p^a)$ :
\bea
v^{-1}R^{(0)}(\x(p)) v u &=& \s(p^a) v u \cr
e_au^T C v^{-1} R^{(0)}(\x(p)) v  &=& \s(p^a)e_a  u^T C
\eea
Indeed we have $Ve_a=\left(\textbf{0},\dots,\textbf{0}, \overset{a}{\textbf{v}_a},\textbf{0},\dots,\textbf{0}\right)$ where $\textbf{0}$ is the $d$-dimensional zero vector and $\textbf{v}_a$ is the $a^{\text{th}}$ eigenvector of $\mathcal{R}^{(0)}(\x(p))$. Consequently, $\mathcal{R}^{(0)}(\x(p))Ve_a=s(p^a)Ve_a$ and inserting $vu=Ve_a$ provides the first identity. The second identity follows from $SV^{-1}=V^{-1}\mathcal{R}^{(0)}$ which is equivalent to $S\mathcal{V}^TCv^{-1}=\mathcal{V}^T C v^{-1}\mathcal{R}^{(0)}$. Multiplying on the left by $e_a=e_a^T$ and observing that $e_a S=S e_a^T=s(p^a)e_a$ and $u^T=e_a^T\mathcal{V}^T$ gives the second identity.
     
Let us denote 
\beq
H = v^{-1} M^{(k),r} v.
\eeq
We must have from \eqref{Polpart} :
\beq\label{eqH1}
H = H u u^T C + u u^T C H=H u u^T C + v^{-1}V e_a u^T C H.
\eeq
multiplying on the right by $u$ gives $Hu = Hu (u^TCu) + u (u^T CHu) = Hu e_a+ u (u^TCHu)$ and thus after a multiplication on the right by $e_a$ :
\beq\label{eqH2}
u^T C H u e_a=0.
\eeq
Moreover, using the fact that $k$ is minimal, the polar part at order $\hbar^k$ of $\hbar \partial_t M = [R,M]$ at $z=p$ implies that
\beq
[H,v^{-1}R^{(0)}(\x(p)) v]=0.
\eeq
This last equation implies that $vHu$ and $e_a u^TCHv^{-1}$ are respectively right and left eigenvectors of $R^{(0)}(\x(p))$ for the eigenvalue $\s(p^a)$. Indeed from the last equation and $u=v^{-1}Ve_a$ :
\beqq v^{-1}\mathcal{R}^{(0)}(\x(p))vHu=Hv^{-1}\mathcal{R}^{(0)}(\x(p))Ve_a=s(p^a)Hv^{-1}V e_a=s(p^a)Hu\eeqq
Similarly, using that $e_a u^T C v^{-1} R^{(0)}(\x(p))= \s(p^a)e_a u^T C v^{-1}$ we get :
\beqq e_au^TCHv^{-1}\mathcal{R}^{(0)}(\x(p))v= e_au^TCv^{-1}\mathcal{R}^{(0)}(\x(p))vH=\s(p^a)e_a u^T C H\eeqq
  
Since $p$ is neither a pole of $L^{(0)}$ (and thus not a pole of $R^{(0)}$) nor a branchpoint, $\s(p^a)$ is not degenerate,  i.e. the eigenspace has dimension one, and therefore, there exist some scalars $\mu,\tilde \mu$ such that
\beq
vHu = \mu u
\qquad , \qquad
e_au^T C Hv^{-1} = \tilde \mu e_a u^T Cv^{-1}.
\eeq
which is equivalent to 
\beq Hu=\mu u \text{  and  } e_au^T C H=\tilde \mu e_a u^T\eeq
In particular, using \eqref{eqH2}, we get
\beq
0=u^TCHue_a=\mu u^tCu e_a = \mu e_a,
\eeq
and thus $\mu=0$ and $Hu=0$. Similarly,
\beq
0 = e_au^TCHue_a = \tilde \mu e_a u^TCue_a= \tilde \mu e_a
\eeq
Hence $\tilde{\mu}=0$ and $e_a u^TCH=0$. Finally we insert the last results into \eqref{eqH1}, this gives $H=0$, and thus $M^{(k),r}=0$, which contradicts our polar assumption. Therefore $M^{(k)}(z)$ has no pole at $z=p$.
}

\subsection{Correlators}
From the matrices $M(x.E,t,\hbar)$, we define the {\it connected} correlators as in \cite{BE09,BBE14} :
\bd[Connected correlators]\label{DefConnectedCorrelators}
We define for $n\geq 1$ the correlators:
\bea
\omega_n(x_1.E_1,\dots,x_n.E_n)= 
\left\{
\begin{array}{lr}
\hbar^{-1} \operatorname{Tr} \left(L(x_1)M(x_1.E_1)\right)dx_1 & n=1 \cr
\frac{(-1)^{n-1}}{n}\underset{\sigma\in\mathfrak S_n}{\sum} \frac{\operatorname{Tr} \underset{i=1}{\overset{n}{\prod}} M(x_{\sigma(i)}.E_{\sigma(i)}) }{\underset{i=1}{\overset{n}{\prod}} (x_{\sigma(i)}-x_{\sigma(i+1)})} \underset{i=1}{\overset{n}{\prod}} dx_i& n\geq 2
\end{array}\right. \cr
\eea
or equivalently the sequence of functions:
\bea
W_n(x_1.E_1,\dots,x_n.E_n)= 
\left\{
\begin{array}{lr}
\hbar^{-1} \operatorname{Tr} \left(L(x_1)M(x_1.E_1)\right) & n=1 \cr
\frac{(-1)^{n-1}}{n}\underset{\sigma\in\mathfrak S_n}{\sum} \frac{\operatorname{Tr} \underset{i=1}{\overset{n}{\prod}} M(x_{\sigma(i)}.E_{\sigma(i)}) }{\underset{i=1}{\overset{n}{\prod}} (x_{\sigma(i)}-x_{\sigma(i+1)})} & n\geq 2
\end{array}\right. \cr
\eea
\ed

These correlators are symmetric $n$-forms on $(\mathbb C\times \Lieg)^n$ and they are linear in each $E_i$. They appear naturally in matrix models and in many enumerative problems \cite{BE09,BBE14,Lie,BookEynard,BDY1,BDY2}. From the CFT point of view, $W_n(x_1.E_1,\dots,x_n.E_n)$ is the correlation function corresponding to the insertions of $n$ currents $J(x_i.E_i)$ \cite{MS15}.

\medskip

Note that like $M(x.E,t,\hbar)$, the $W_n$'s are defined as $\hbar$-formal series. We will describe more precisely the $\hbar$ expansions and their properties below. We will also need the non-connected version of the correlators:

\bd[Non-connected correlators]\label{DefCorrelators} The correlators (non connected) are defined from the connected ones by summing over partitions. Denoting $X_i=x_i.E_i$ we define :
\bea
\td{\omega}_n(X_1,\dots,X_n;t,\hbar)&=& \sum_{\mu \,\vdash \{X_1,\dots,X_n\}}\prod_{i=1}^{\ell(\mu)}  \omega_{|\mu_i|}(\mu_i;t,\hbar)\cr
\td{W}_n(X_1,\dots,X_n;t,\hbar)&=& \sum_{\mu \,\vdash \{X_1,\dots,X_n\}}\prod_{i=1}^{\ell(\mu)}  W_{|\mu_i|}(\mu_i;t,\hbar)
\eea
where we sum over all partitions of the set $\{X_1,\dots,X_n\}$ of $n$ points.
For example
\beq
\td{\omega}_1(X_1;t) = \omega_1(X_1;t),
\eeq
\beq
\td{\omega}_2(X_1,X_2;t)
= \omega_1(X_1;t) \omega_1(X_2;t) + \omega_2(X_1,X_2;t)
\eeq
\bea
\td{\omega}_3(X_1,X_2,X_3;t)
&=& \omega_1(X_1;t) \omega_1(X_2;t) \omega_1(X_3;t) + \omega_1(X_1;t) \omega_2(X_2,X_3;t) \cr
&& + \omega_1(X_2;t) \omega_2(X_1,X_3;t) + \omega_1(X_3;t) \omega_2(X_1,X_2;t) \cr
&& + \omega_3(X_1,X_2,X_3;t) 
\eea
and so on...
\ed

\br
One often says that the connected correlators are the ``cumulants'' of the non-connected ones.
\er

\subsection{Tau function}

We also recall for bookkeeping (indeed we shall not use it in this article) the definition of the Tau-function by Miwa-Jimbo \cite{JMI,JMII}.
Let $T(x)={\rm diag}(T_1(x),\dots,T_i(x))$ (it is the exponential term of the WKB expansion \eqref{WKBpsi}), such that
\beq
\frac{\partial T_i(x)}{\partial x} = y_i(x)
\quad , \quad
\frac{\partial T_i(x)}{\partial t} = s_i(x),
\eeq
In other words :
\beq
\frac{\partial T(x)}{\partial t} = \sum_{i=1}^d s_i(x) e_i
\qquad \text{with} \quad
e_i=\text{diag}(0,\dots,0,\overset{i}{1},0,\dots,0).
\eeq
The Miwa-Jimbo-Ueno-Takasaki definition of the Tau function is \cite{JMI,JMII}
\beq
\hbar \frac{\partial \ln\Tau}{\partial t} 
=  \sum_{q={\rm poles\,of\,}\x,\s} \Res_{x\to q} \operatorname{Tr} \left(\frac{\partial T(x)}{\partial t} \Psi(x)^{-1} \frac{\partial\Psi(x)}{\partial x}\right)\,\,dx .
\eeq
Let us rewrite it in our notations, using $W_1$ :
\bea
\hbar \frac{\partial \ln\Tau}{\partial t} 
&=&  \sum_{q={\rm poles\,of\,}\x,\s} \Res_{x\to q} \operatorname{Tr} \left(\frac{\partial T(x)}{\partial t} \Psi(x)^{-1} \frac{\partial\Psi(x)}{\partial x}\right)\,\,dx \cr
&=&  \hbar^{-1} \sum_{q={\rm poles\,of\,}\x,\s} \underset{i=1}{\overset{d}{\sum}} \Res_{x\to q} s_i(x) \operatorname{Tr} \left(e_i \Psi(x)^{-1} L(x) \Psi(x)\right)\,\,dx \cr
&=&  \hbar^{-1} \sum_{q={\rm poles\,of\,}\x,\s} \underset{i=1}{\overset{d}{\sum}} \Res_{x\to q} s_i(x) \operatorname{Tr}  \left(L(x) \Psi(x)e_i \Psi(x)^{-1}\right)\,\,dx \cr
&=&  \sum_{q={\rm poles\,of\,}\x,\s} \underset{i=1}{\overset{d}{\sum}} \Res_{x\to q}  s_{i}(x) \omega_1(x.e_i).
\eea
In particular it explains why the one-form $\omega_1(x.E)$ is so useful. For $n\geq 2$, the notation $W_n$ follows the definition of correlation functions arising in topological recursion and in the study of random Hermitian matrices. In fact it has been shown recently in a series of papers \cite{BE09,BBE14} that under suitable conditions, known as Topological Type property, the correlation functions presented in definition \ref{DefCorrelators} can be reconstructed from the application of the topological recursion to the spectral curve $E(x,y)=0$ attached to the differential system. The precise statement can be found in \cite{BBE14} and will be summarized in the next section for our purposes.

\section{Topological Type property : definition and proof\label{S4}}

\subsection{Topological Type property}

We recall the definition $3.3$ of \cite{BBE14}, specialized to the case of a genus $0$ spectral curve (thus skipping many unnecessary geometric technicalities appearing only when the genus is strictly positive) :

\begin{proposition}[Definition $3.3$ of \cite{BBE14}] 
A sequence of differential forms $(\omega_n)_{n\geq 1}$ (or equivalently functions $(W_n)_{n\geq 1}$) is said to have an expansion of topological type (TT property) when :
\begin{enumerate}
\item\label{tt1} \underline{Existence of an expansion in $\hbar$} : The $\omega_n$'s and $W_n$'s are formal series of $\hbar$ of the form:
\bea \omega_n(X_1,\dots,X_n;t,\hbar)&=&\sum_{k=-\delta_{n,1}}^{\infty} \omega_n^{(k)}(X_1,\dots,X_n;t)\,\hbar^k\cr
\Leftrightarrow \quad W_n(X_1,\dots,X_n;t,\hbar)&=&\sum_{k=-\delta_{n,1}}^{\infty} W_n^{(k)}(X_1,\dots,X_n;t)\,\hbar^k
\eea
whose coefficients, denoted
\beq
\omega_n^{(k)}(z^{a_1}(x_1),\dots,z^{a_n}(x_n))=W_n^{(k)}(x_1.e_{a_1},\dots,x_n.e_{a_n};t)dx_1\dots dx_n
\eeq
are rational functions of their arguments (this is where we use that the genus of the curve is zero). Moreover, the one-form $\omega_1^{(0)}(z)$ is required to be the Liouville form :
\beq
\omega_1^{(0)}(z) = \y(z)\x'(z)dz = \y(z)d\x(z).
\eeq
and the bi-differential form $\omega_2^{(0)}(z_1,z_2)$ is required to correspond to the fundamental $2^{\text{nd}}$ kind differential (``Bergmann-Schiffer-Klein kernel'') of the Riemann sphere (also specific to genus $0$ curve) :
\beq
\omega_2^{(0)}(z_1,z_2) = B(z_1,z_2) = \frac{dz_1dz_2}{(z_1-z_2)^2}.
\eeq
\item \label{ttloopeq} \underline{Loop equations} : The $W_n$'s satisfy loop equations, i.e.
for all $1\leq k \leq d$, for all $n\geq 0$ and for all $X_1,\dots,X_n$ with $X_i=x_i.E_i$ ($E_i\in \Lieh$) the following quantity 
\beq
\sum_{1\leq i_1<i_2<\dots<i_k\leq d} \tilde W_{k+n}(x.e_{i_1},\dots,x.e_{i_k},X_1,\dots,X_n)
= \tilde P_{k,n}(x;X_1,\dots,X_n)
\eeq
is well defined when $x$ and all $x_i$ are distinct. Moreover it is a rational function of $x$ such that the meromorphic one-form of $z$ :
\beqq
\eta_n(z;X_1,\dots,X_n)=
\left(\frac{y^d + \underset{k=1}{\overset{d}{\sum}} (-1)^k y^{d-k} \tilde P_{k,n}(x;X_1,\dots,X_n)}{E_y(x,y)}\,dx\right)_{x=\x(z),y=\y(z)}
\eeqq
has no pole at the poles of $\x$ and $\y$ neither at branchpoints. Its only poles may be at double points (zeros of $E_y(\x,\y)$) and/or at coinciding points $\x(z)=x_i$ for some $i$.
\item\label{tt2} \underline{Pole property} : For $(n,k)\notin\{(1,-1),(2,0)\}$, the rational differential forms $\omega_n^{(k)}(z_1,\dots,z_n)$ may only have poles at the branchpoints of the spectral curve. In particular they must have no pole at double points of the spectral curve, nor at coinciding points $z_i=z_j$ with $i\neq j$.
\item\label{tt3} \underline{Parity property} : Under the change $\hbar \leftrightarrow -\hbar$ the correlation functions satisfy $(\omega_n)_{-\hbar}=(-1)^n (\omega_n)_\hbar$. This is equivalent to say that the $\hbar$-expansions of the $\omega_n$ only contain even (resp. odd) exponents in $\hbar$ when $n$ is even (resp. odd).
\item\label{tt4} \underline{Leading order property} : For $n\geq 1$ we have $\omega_n=O\left(\hbar^{n-2}\right)$. 
\end{enumerate}
Note that combining the existence of an $\hbar$-expansion, the parity property and the leading order property is equivalent to say that : 
\bea 
\forall \, n\geq 1 \, :\, W_n&=&\sum_{g=0}^\infty \hbar^{2g-2+n} W_{g,n}\quad , \quad W_{g,n} = W_n^{(n+2g-2)}\cr
\Leftrightarrow \quad \omega_n&=&\sum_{g=0}^\infty \hbar^{2g-2+n} \omega_{g,n}\quad , \quad \omega_{g,n} = \omega_n^{(n+2g-2)}.
\eea 
\end{proposition}

It was proved in \cite{Lie} that the TT property and the loop equations (here obtained by construction for the $(W_n)_{n\geq 1}$'s \cite{BBE14}) imply that the coefficients $\left(\omega_{n}^{(g)}\right)_{n\geq 1,g\geq 0}$ satisfy the topological recursion of \cite{EOFg}.

We now claim that our assumptions \ref{Asshbarexpansion}, \ref{genuszeroassumption}, \ref{Assnodblpt}, \ref{AssumptionV}, \ref{Asspolesordrek} and \ref{Asssym} imply that the Topological Type property is satisfied for the set of correlators $(\omega_n)_{n\geq 1}$. Notice that assumption \ref{Asshbarexpansion} is usually automatically verified in the formal context while assumptions \ref{genuszeroassumption}, \ref{Assnodblpt} and \ref{AssumptionV} can be verified from only $\hbar^0$ computations. Eventually only assumptions \ref{Asspolesordrek} and \ref{Asssym} require general properties (location and number of poles, Hamiltonian structure, etc.) of the Lax system.

\subsection{Proof of condition \ref{tt1} of the TT property: Existence of an $\hbar$ expansion for correlators}

The existence of an $\hbar$ expansion for the correlation functions is an immediate corollary of theorem \ref{Mexp} for $M(x.E,t,\hbar)$. Indeed, inserting the series expansion of $M(x.E,t)$ in definition \ref{DefConnectedCorrelators} provide the wanted $\hbar$ expansion, whose coefficients are indeed rational functions of the $z^i(x)$. Therefore only the explicit computations of $\omega_{1}^{(0)}(z)$ and $\omega_2^{(0)}(z_1,z_2)$ remain to prove.

- The computation of $\omega_{1}^{(0)}(z)$ is straightforward from the definition : 
\bea
W_{1}^{(0)}(x.e_a) &=& \Tr M^{(0)}(x.e_a)L^{(0)}(x) = \Tr V(x) e_a V(x)^{-1} V(x) Y(x) V(x)^{-1} \cr
&=& \Tr e_a Y(x)= Y_a(x)= \y(z^a(x)).
\eea
Eventually it gives :
\beq
\omega_1^{(0)}(z) = \y(z)\x'(z) dz.
\eeq

- The computation of $\omega_2^{(0)}(z_1,z_2)$ is a direct consequence of corollary \ref{corB}. Hence condition \ref{tt1} of the TT property is proved.

\subsection{Proof of condition \ref{ttloopeq} of the TT property : Loop equations}

The proof is already done in \cite{BEO13}, and we shall also use a rewriting as in \cite{BouchardEynard2016}.

\subsubsection{Rewriting of the loop equations}

Recall that $e_1,\dots,e_d$ span a Cartan subalgebra $\Lieh$ when the Lie group is taken to be $G=GL_d(\mathbb C)$. Using this basis, the Casimirs of $\Lieg=\mathfrak{gl}_d(\mathbb{C})$ are
\beq
C_k = \sum_{1\leq i_1<i_2<\dots<i_k\leq d} e_{i_1} \otimes \dots \otimes e_{i_k}.
\eeq
Consequently, the loop equations derived in \cite{BEO13} are :

\begin{theorem}[From \cite{BEO13}]\label{thLoopeq1}
If $L(x)$ is a rational function of $x$, then the non-connected correlators $\tilde W_n$'s  are such that for all $1\leq k \leq d$, for all $n\geq 0$, and for all $X_1,\dots,X_n$ with $X_i=x_i.E_i$ where $E_i\in \Lieh$, the following quantity : 
\beq \label{Resultat}
\sum_{1\leq i_1<i_2<\dots<i_k\leq d} \tilde W_{k+n}(x.e_{i_1},\dots,x.e_{i_k},X_1,\dots,X_n)
= \tilde P_{k,n}(x;X_1,\dots,X_n)
\eeq
is well defined for $x$ and all $x_i$ distinct. Moreover, it is a rational function of $x$, with only possible poles at the poles as $L(x)$ and at coinciding points $x=x_i$ for some $i$.
\end{theorem}

Saying that it is well defined is not obvious, because $\tilde W_n$ has poles at coinciding points, due to the presence of $W_2$ factors. However, $W_2(x.e_{i_1},x'.e_{i_2})$ has no pole on the diagonal $x=x'$ if $i_1\neq i_2$, and the summation in \eqref{Resultat} is only on distinct indices $i_1\neq i_2 \neq \dots \neq i_k$.

\medskip

Another version of the same theorem can be written after partially decomposing the non-connected correlators $\tilde W_{k+n}$'s into the connected ones $W_{i}$. It is given by:

\bc\label{corLoopeq1}
For all $1\leq k \leq d$, for all $x$, for all $n\geq 0$, and for all $X_i=x_i.E_i$ (with $x$ and all $x_i$'s are distinct), the following quantity :
\beq
\sum_{1\leq i_1<i_2<\dots<i_k\leq d} {\mathcal W}_{k,n}(x.e_{i_1},\dots,x.e_{i_k},X_1,\dots,X_n)
= P_{k;n}(x;X_1,\dots,X_n)
\eeq
is a rational function of $x$. In this formula, ${\mathcal W}_{k,n}(x.e_{i_1},\dots,x.e_{i_k},X_1,\dots,X_n)$ stands for 
\beq
 {\mathcal W}_{k,n}(K;A)
= \sum_\ell \sum_{(I_1,\dots,I_\ell)\,\vdash\, K\,\, ; \,\, J_1\uplus J_2 \uplus \dots \uplus J_\ell=A} \prod_{i=1}^{\ell} W_{|I_i|+|J_i|}(I_i \cup J_i)
\eeq
where we denoted the ensembles $A=\{X_1,\dots,X_n\}$ and $K=\{x.e_{i_1},\dots,x.e_{i_k}\}$. In other words, we sum over all partitions of $K$ into non-empty parts, and we associate to each part $I_i$ of $K$ a (possibly empty) part $J_i$ of $A$, in all possible ways.
\ec

In short, every $I_i\cup J_i$, which is a part of $K\cup A$, contains at least one element of $K$ but $J_i$ may be empty. This is nearly the same definition as $\tilde W_{k+n}$, but the latter may contain parts without elements of $K$ (i.e. $I_i=\emptyset$). For example, we have:
\beq
{\mathcal W}_{2,1}(X,X';X_1) = W_1(X) W_2(X',X_1) + W_1(X')W_2(X,X_1)+W_2(X,X',X_1)
\eeq
which differs from 
\bea
\tilde W_3(X,X',X_1)
&=& W_1(X) W_2(X',X_1) + W_1(X')W_2(X,X_1)+W_2(X,X',X_1) \cr
&& + W_1(X_1)\,\left(W_1(X)W_1(X')+W_2(X,X')\right) \cr
&=& {\mathcal W}_{2,1}(X,X';X_1) + W_1(X_1)\,{\mathcal W}_{2,0}(X,X').
\eea

Let us reformulate again the loop equations by summing over $k$ and making a generating series with a formal variable $y$ :

\bc The following quantity :
\bea
&&Q_{n}(x,y;X_1,\dots,X_n)= y^d \delta_{n,0} \cr
&& + \sum_{k=1}^d (-1)^k \hbar^k y^{d-k} \sum_{1\leq i_1<\dots <i_k\leq d}
{\mathcal W}_{k,n}(x.e_{i_1},\dots,x.e_{i_k},X_1,\dots,X_n)
\eea
is a polynomial of $y$ of degree $\leq d$. It is also a rational function of $x$ with only possible poles at the poles of $L(x)$ and at coinciding points $x=x_i$.
\ec

Eventually, it is useful to separate the leading term of the $1$-point function $W_1(x.e_a)=\hbar^{-1}\,\y(z^a(x)) + O(\hbar)$ from the subleading ones. We thus introduce the following  :

\bd\label{defWnhatcal}
We define :
\bea
\hat W_1(x.e_a) &=& W_1(x.e_a) - \hbar^{-1}\,\y(z^a(x))  \text{  for  } n=1\cr
\hat W_n&=&W_n  \text{  for  } n>1.
\eea
Then we define the hat-disconnected correlators :
\beq
\hat {\tilde W}_{n}(A)
= \sum_\ell \sum_{(I_1,\dots,I_\ell)\,\vdash\, A} \prod_{i=1}^{\ell} \hat W_{|I_i|}(I_i),
\eeq
and the hat-partially disconnected correlators :
\beq
\hat \W_{k,n}(K;A)
= \sum_\ell \sum_{(I_1,\dots,I_\ell)\,\vdash\, K\,\, ; \,\, J_1\uplus J_2 \uplus \dots \uplus J_\ell=A} \prod_{i=1}^{\ell} \hat W_{|I_i|+|J_i|}(I_i \cup J_i)
\eeq
In other words, we have the same definition as $\tilde W_n$ and $\W_{k,n}$ respectively but with the factors $W_j$'s replaced by $\hat W_j$'s.
\ed
\br \label{remarktildeW}
Notice that $\hat {\tilde W}_{k+n}$ and  $\hat \W_{k,n}$ are related :
\beq
\hat {\td W}_{k+n}(K\cup A) = \sum_{A'\subset A} \hat {\W}_{k,|A'|}(K;A') \hat {\td W}_{|A|-|A'|}(A\setminus A').
\eeq
i.e. $\hat{\td W}_n$'s are linear combinations of $\hat\W_{k,n}$'s with coefficients independent of $K$. This is particularly convenient since it means that every statement about the analytic structure of the $\hat{\td W}_n$'s is immediately transmitted to $\hat\W_{k,n}$ and vice-versa.
\er

\medskip

Eventually, the loop equations can be reformulated (see \cite{BouchardEynard2016}) in another way :

\bc[Loop equation, version of \cite{BouchardEynard2016}]\label{CorrolaryLoop}
The following quantities :
\bea
\tilde P_{n}(x,y;X_1,\dots,X_n)
&=& \sum_{I\subset \{x.{e_1},\dots,x.{e_d}\}} \hbar^{|I|}\,
 \hat{\tilde W}_{|I|,n}(I;X_1,\dots,X_n) \,\prod_{a\notin I} (y-\y(z^a(x))) \cr
&=& E(x,y)\, \sum_{I\subset \{x.{e_1},\dots,x.{e_d}\}} \hbar^{|I|}\,
\frac{ \hat{\td W}_{|I|,n}(I;X_1,\dots,X_n)}{\prod_{a\in I} (y-\y(z^a(x)))} ,
\eea
and
\bea
P_{n}(x,y;X_1,\dots,X_n)
&=& \sum_{I\subset \{x.{e_1},\dots,x.{e_d}\}} \hbar^{|I|}\,
 \hat\W_{|I|,n}(I;X_1,\dots,X_n) \,\prod_{a\notin I} (y-\y(z^a(x))) \cr
&=& E(x,y)\, \sum_{I\subset \{x.{e_1},\dots,x.{e_d}\}} \hbar^{|I|}\,
\frac{ \hat\W_{|I|,n}(I;X_1,\dots,X_n)}{\prod_{a\in I} (y-\y(z^a(x)))}
\eea
are polynomials of $y$ of degree $\leq d$. Moreover they are rational functions of $x$ with only possible poles at the poles of $L(x)$ and at coinciding points $x=x_i$.
\ec

In fact in \cite{Lie} the explicit expression of $\tilde P_n(x,y;X_1,\dots,X_n)$ was derived :

\bt[From \cite{Lie}]\label{thPnprodM}We have :
\beq \label{LoopLoop}
\tilde P_n(x,y;X_1,\dots,X_n)
= [\epsilon_1\epsilon_2 \dots \epsilon_n] \,\det\left(y {\rm Id}_d - L(x) - F_{\boldsymbol{\epsilon}}(X_1,\dots,X_n) \right)
\eeq
where
\bea
F_{\boldsymbol{\epsilon}}(X_1,\dots,X_n)
&=& \sum_{i=1}^n \frac{\epsilon_i M(X_i)}{(x-x_i)(x_i-x)} + \sum_{i\neq j}\frac{\epsilon_i\epsilon_j M(X_i)M(X_j)}{(x-x_i)(x_i-x_j)(x_j-x)} \cr
&& + \sum_{k=3}^{n} \sum_{i_1\neq i_2\neq \dots \neq i_k} \frac{\epsilon_{i_1}\dots \epsilon_{i_k} M(X_{i_1})\dots M(X_{i_k})}{(x-x_{i_1})(x_{i_1}-x_{i_2})\dots (x_{i_k}-x)}
\eea
and where the notation is such that $[\epsilon_1^{k_1}\epsilon_2^{k_2} \dots \epsilon_n^{k_n}] f(\boldsymbol{\epsilon})$ is the coefficient of $\epsilon_1^{k_1}\epsilon_2^{k_2} \dots \epsilon_n^{k_n}$ of the polynomial $f(\boldsymbol{\epsilon})$.

\et

\subsubsection{$\hbar$ expansion of the loop equations}

Since the right hand side of the loop equations \eqref{LoopLoop} has an $\hbar$ expansion (from assumption \ref{Asshbarexpansion} and theorem \ref{Mexp}), so does the left hand side. Thus we write :
\bea
P_{n}(x,y;X_1,\dots,X_n) &=& \sum_{j\geq 0} \hbar^j P^{(j)}_{n}(x,y;X_1,\dots,X_n)\cr
\tilde P_{n}(x,y;X_1,\dots,X_n) &=& \sum_{j\geq 0} \hbar^j P^{(j)}_{n}(x,y;X_1,\dots,X_n).
\eea
where each $P^{(j)}_{n}(x,y;X_1,\dots,X_n)$ and $\tilde P^{(j)}_{n}(x,y;X_1,\dots,X_n)$ is a polynomial of $y$ of degree $\leq d$ and a rational function of $x$.

\subsubsection{Specializing the loop equations on the spectral curve}

Let us denote :
\beq
D=\{x.e_1,\dots,x.e_d\}\quad,\quad A=\{X_1,\dots,X_n\}.
\eeq
We isolate terms $I=\emptyset$ and $|I|=1$ in Corollary \eqref{CorrolaryLoop}. We find :
\bea P_n(x,y;A)&=&\hat{\W}_{n}(A) E(x,y)+ \sum_{i=1}^d \hbar \hat{\W}_{1,n}(x.e_{i};A)\prod_{j\neq i} (y-\y(z^{j}(x))) \quad \quad \quad\cr
&&+\sum_{I\subset D \,/\,|I|\geq 2} \hbar^{|I|}\hat{\W}_{|I|,n}(I;A)\frac{E(x,y)}{\underset{i\in I}{\prod}(y-\y(z^{i}(x)))}
\eea
Specializing at $y=\y(z^{i_0}(x))$ for a given $i_0$ we observe that the first term vanishes while the second term only restricts to $i=i_0$. We find :
\bea 
&&P_n(x,\y(z^{i_0}(x));A)=\hbar\hat{\W}_{1,n}(x.e_{i_0};A) E_y(x,\y(z^{i_0}(x))) \cr
&&+\sum_{\{i_0\}\subset I\subset D \,/\,|I|\geq 2}\hbar^{|I|} \hat{\W}_{|I|,n}(I;A)\prod_{i\notin I}(\y(z^{i_0}(x))-\y(z^{i}(x))).
\eea
We recognize here $E_y(x,y)=\partial_y E(x,y)=\underset{i=1}{\overset{d}{\sum}}\underset{j\neq i}{\prod}(y-\y(z^j(x)))$ evaluated at the point $y=\y(z^{i_0}(x))$. Moreover $\hat{\W}_{1,n}(x.E_{i_0};A)=W_{n+1}(x.e_{i_0},A)$. 
Indeed, in the definition of $\hat{\W}_{1,n}(x.e_{i_0};A)$, all parts must contain $x.e_{i_0}$ and thus there must be exactly one part.
In the end, we have :
\bea \label{Core} &&P_n(x,\y(z^{i_0}(x));A)=\hbar W_{n+1}(x.E_{i_0},A)E_y(x,y(z^{i_0}(x))) \quad \qquad\cr
&&+\sum_{\{i_0\}\subset I\subset D \,/\,|I|\geq 2}\hbar^{|I|} \hat{\W}_{|I|,n}(I;A)\prod_{i\notin I}(\y(z^{i_0}(x))-\y(z^{i}(x)))
\eea

\subsubsection{Poles of $P_n^{(k)}$}

Theorem \ref{thPnprodM} together with assumption \ref{Asspolesordrek} imply :
 
\bc\label{Pnkpoles}
For every $(k,n)\in\mathbb{N}^2\setminus \{(0,0)\}$ and for every generic $X_1,\dots ,X_n$, the function
\beq
P^{(k)}_n(\x(z),\y(z);X_1,\dots,X_n) \frac{d\x(z)}{E_y(\x(z),\y(z))}
\eeq
is a one-form on the Riemann sphere, whose poles may only be at coinciding points or at double points. Thanks to remark \ref{remarktildeW}, the same applies to $\tilde P_n^{(k)}$.
\ec

\subsection{Proof of condition \ref{tt2} of the TT property : The pole structure \label{SectionPole}}
We want to prove that for $(k,n)\notin\{(-1,1),(0,2)\}$, the only poles of $\omega^{(k)}_{n}(z,z_2,\dots,z_n)$ may be at branchpoints. By definition, the only possible singularities may arise at branchpoints, double points, coinciding points and punctures (i.e. simple poles of $\x$ or poles of $\y$).

\begin{itemize}
\item \textit{No poles at double points} : We have proved that the $\hbar^k$ term $M^{(k)}$ is a rational function of $z^a(x)$ without poles at double points, so by definition all $W_n$ cannot have poles at double points. This implies that the $\omega_n^{(g)}$ are regular at the double poles.
\item \textit{No poles at coinciding points} : By definition, the $W_n$'s involve denominators $\frac{1}{x_i-x_j}$ that may lead to poles at coinciding points.
However, for $n>2$, the poles at coinciding points may be at most simple poles and the residue is a sum over permutations, that contains both each permutation and its inverse having opposite residues. Therefore the total residue vanishes and there is no pole at coinciding points. For $n=2$, the pole at coinciding points may be a double pole. More precisely, the coefficient of the double pole is 
\beq
\lim_{x_2\to x_1} \Tr \left(M(x_1.e_a)M(x_2.e_b)\right) = \Tr (e_a e_b )= \delta_{a,b}.
\eeq
which is independent of $\hbar$ and thus $W_2^{(k)}$ has no double pole for $k>0$. Eventually, there might be a simple pole in $W_2^{(k)}(x_1,x_2)$ at $x_1=x_2$, but the symmetry $W_2^{(k)}(x_1,x_2)=W_2^{(k)}(x_2,x_1)$ implies that the residue must vanish. Therefore $W_2^{(k)}(x_1,x_2)$ has no pole at coinciding points, for $k>0$. Consequently all differentials $\omega_n^{(k)}$ with $(k,n)\notin\{(-1,1),(0,2)\}$ are regular at coinciding points.
\item \textit{No poles at punctures (i.e. simple poles of $\x$ or $\y$)} :
In principle, $M^{(k)}(z)$ may have poles at poles of $\x$ and $\y$ (poles of $L^{(0)}$), so the $\omega^{(k)}_n$ may also have such poles. We shall prove by induction on $k+n$ that for $k+n\geq 0$, $\omega^{(k)}_n$ has no pole at the punctures.

- This is clearly true for $\omega^{(0)}_2$ from Corollary \ref{corB}.

- This is also true for $\omega_1^{(0)}$. Indeed from \eqref{Core} with $n=0$ and $k=1$ we get :
\beq
\omega^{(0)}_1(z) = \frac{P_0^{(1)}(\x(z),\y(z))\,d\x(z)}{E_y(\x(z),\y(z))}
\eeq
From corollary \ref{Pnkpoles}, the right hand side cannot have poles at the poles of $\x$ or $\y$. Note that this implies that $\omega^{(0)}_1(z)$ has no pole at all, and therefore $\omega^{(0)}_1(z)=0$.

- Let us assume that $\omega^{(k')}_{n'}$ have no pole at poles of $\x$ or $\y$ for all $k'+n'<k+n$. Writing \eqref{Core} with $A=\{X_2,\dots,X_n\}$ we get :
\beaa
\omega^{(k)}_n(z;A)
&=& \frac{P_{n-1}^{(k)}(\x(z),\y(z);A)\,d\x(z)}{E_y(\x(z),\y(z))} \cr
&&  -  \sum_{\{i_0\}\subset I\subset D \,/\,|I|\geq 2}  \frac{\hat{\W}^{(k-|I|)}_{|I|,n-1}(I;A) (d\x(z))^{|I|}}{\prod_{i\in I\setminus\{i_0\}}(\y(z^{i_0}(x))-\y(z^{i}(x)))d\x(z)}
\eeaa
The term on the first line has no pole at the punctures from corollary \ref{Pnkpoles}. The numerator on the second line only involves $k-|I|+|I|+n-1<k+n$, and so by induction hypothesis, the numerator has no pole at punctures.
The denominator also does not vanish at the punctures (notice that it vanishes only at branchpoints and double points). Therefore we prove the property for $k+n$ and we conclude by induction.
\end{itemize}

\subsection{Proof of condition \ref{tt3} of the TT property : The parity property}

It was proved in \cite{BBE14} that assumption \ref{Asssym} is a sufficient condition to get the parity property and we shall not redo the (easy) proof of \cite{BBE14} here. We just mention that we do not know if the converse is true : is assumption \ref{Asssym} also a necessary condition to get the parity property? At the moment we do not know any counter-example and all known integrable systems that we have been looked at satisfy assumption \ref{Asssym}.

\subsection{Proof of condition \ref{tt4} of the TT property : The leading order property}

This condition is the hardest to prove. In \cite{BE09,BBE14} a method called ``loop insertion operator'' was used, and part of the proof was missing (this has been fixed for instance in chapter 5 of \cite{BookEynard}, for the Painlev\'e I hierarchy, i.e. $(p,2)$ minimal models). We shall not pursue here this complicated method.

Instead, in \cite{IwakiMarchal}, another proof, based on loop equations, was presented, but only for $2\times 2$ systems. We shall extend this loop equation method to higher rank systems. The generalization is not straightforward, because loop equations are much more involved in higher rank. It is obvious that the proof can be done for $W_n$ or $\omega_n$ equivalently since they are related by a multiplication by $dx_1\dots dx_n$ that does not depend on $\hbar$.

We shall prove by induction on $k\geq 1$ that :
\bt \label{TheoInduction}
The following proposition $\mathcal{P}_k$ holds for $k\geq 1$ :
\begin{center} $\mathcal{P}_k$ : For all $j\geq k$ : $W_j=O(\hbar^{k-2})$.
\end{center}
\et

\textbf{Proof :} We first observe that $\mathcal{P}_1$ and $\mathcal{P}_2$ are trivially verified. Indeed, by definition (see the $\hbar$ expansion of $M(x.E)$ in theorem \ref{Mexp}) $W_1(x_1.E_{1})$ is of order $\hbar^{-1}$ while all other correlation functions $W_n(x_1.E_{1},\dots,x_n.E_{n})$ with $n\geq 2$ are at least of order $\hbar^0$.

Let us now assume that for a given $n\geq 2$, propositions $\mathcal{P}_1$ up to $\mathcal{P}_n$ are verified. We now need to control the order of the last term of \eqref{Core} and thus of 
\beqq \hat{\W}_{|I|,n}(I;A)=\underset{(I_1,\dots, I_l)\vdash I\,\, ; \,\, A_1\sqcup\dots\sqcup A_l=A}\sum \quad \underset{i=1}{\overset{l}{\prod}} \hat{W}_{|I_i|+|A_i|}(I_i,A_i)\eeqq
where we recall that none of the parts $I_i$ are allowed to be empty.

There are $3$ different cases :
\begin{enumerate}\item $|I_i|=1$ and $A_i=\emptyset$. In that case we get $\hat{W}_{1}(x.e_i)$ which is of order at least $\hbar^0$ (because we have a hat version of $W_1$ whose $\hbar^{-1}$ term is removed).
\item $1<|I_i|+|A_i|\leq n$. In that case we can apply $\mathcal{P}_{|I_i|+|A_i|}$ and thus we get an order of $\hbar^{|I_i|+|A_i|-2}$.
\item $|I_i|+|A_i|>n$. In that case, we can only apply $\mathcal{P}_n$ and thus we get an order of $\hbar^{n-2}$.
\end{enumerate} 
Consequently we get :
\beq \hat{W}_{|I_i|+|A_i|}(I_i,A_i)=O\left(\hbar^{\text{Min}(n,|I_i|+|A_i|)-2+\delta_{|I_i|+|A_i|=1}}\right)\eeq
Thus, we obtain a term of order :
\beq \label{OrdreEnCours}\hbar^{|I|}\prod_{i=1}^l \hat{W}_{|I_i|+|A_i|}(I_i,A_i)=O\left(\hbar^{\underset{i=1}{\overset{l}{\sum}}\left(\text{Min}(n,|I_i|+|A_i|)-2+\delta_{|I_i|+|A_i|=1}\right)+|I| }\right)\eeq
We shall prove the following inequality :
\beq \label{Inequality} \underset{i=1}{\overset{l}{\sum}}\left(\text{Min}(n,|I_i|+|A_i|)-2+\delta_{|I_i|+|A_i|=1}\right)+|I|-n\geq 0\eeq
with $\underset{i=1}{\overset{l}{\sum}}|A_i|=n$ ,$|I_i|\geq 1$ and $\underset{i=1}{\overset{l}{\sum}} |I_i|=|I|$. In the case when $l=1$, we have $|A_1|=n$ and $|I_1|=|I|\geq 1$ so we end up with $n-2+1+|I|-n=|I|-1\geq 0$ since $I$ cannot be empty. 
For our future discussion, we will denote $L_1$ the set of indexes for which $1<|I_i|+|A_i|\leq n$ and $L_2$ the set of indexes for which $|I_i|+|A_i|>n$. We will denote respectively $l_1=|L_1|$ and $l_2=|L_2|$ satisfying $l_1+l_2=l$. The case where $l_2=0$ i.e. the minimum is always equal to $|I_i|+|A_i|$ is trivial since in that case we end up at least with $|I|+n-2l+|I|-n=2(|I|-l)\geq 0$ since all $I_i$ have at least one element $|I_i|\geq 1$. 

Let us now consider the general case where $l\geq 2$ and $l_2\geq 1$. We first observe :
\bea \label{UsefulIdentity}|I|&=&\sum_{i\in L_1}|I_i|+\sum_{i\in L_2}|I_i|\cr
&\geq& l_1+\sum_{i\in L_2} |I_i|=l_1+\sum_{i\in L_2}(|I_i|+|A_i|)-\sum_{i\in I_2}|A_i|\cr
&\geq& l_1+l_2(n+1)-\sum_{i \in L_2}|A_i|\cr
&\geq& l_1+l_2(n+1)-n
\eea
Inserting \eqref{UsefulIdentity} into \eqref{Inequality} we obtain :
\bea &&\underset{i=1}{\overset{l}{\sum}}\left(\text{Min}(n,|I_i|+|A_i|)-2+\delta_{|I_i|+|A_i|=1}\right)+|I|-n\cr
&&=\underset{i\in L_1}{\sum}\left(|I_i|+|A_i|-2+\delta_{|I_i|+|A_i|=1}\right)+(n-2)l_2+|I|-n\cr
&&\geq \underset{i\in L_1}{\sum}\left(|I_i|+|A_i|-2+\delta_{|I_i|+|A_i|=1}\right)+(n-2)l_2+l_1+l_2(n+1)-n-n\cr
&&=\underset{i\in L_1}{\sum}\left(|I_i|+|A_i|-2+\delta_{|I_i|+|A_i|=1}\right)+2(nl_2-n-l_2)+l_1+l_2\cr 
&&=\underset{i\in L_1}{\sum}\left(|I_i|+|A_i|-2+\delta_{|I_i|+|A_i|=1}\right)+2(n-1)(l_2-1)+l-2
\eea
The terms in the first sum are non-negative. Then, since $n\geq 2$, and $l_2\geq 1$, $(n-1)(l_2-1)$ is always non-negative. Then, since $l\geq 2$, the last term is also non-negative, thus concluding the proof of inequality \eqref{Inequality}.

Going back to \eqref{OrdreEnCours} and inserting inequality \eqref{Inequality}, we obtain that the second line of \eqref{Core} is at least of order $O(\hbar^{n})$. Consequently, for any $k>0$, evaluating at order $\hbar^{n-k}$ in \eqref{Core} provides:
\beq P_n^{(n-k-1)}(x,y^{i_0}(x);A)=W_{n+1}^{(n-k)}(x.e_{i_0},A) E_y(x,\y(z^{i_0}(x)))\eeq
From $\mathcal P_n$ we know that the right hand side vanishes for $k>1$. Therefore the only possibly non-vanishing term is
\beq \label{Contradiction}W_{n+1}^{(n-2)}(x.e_{i_0},A)dx=\frac{P_n^{(n-1)}(x,\y(z^{i_0}(x));A)}{E_y(x,\y(z^{i_0}(x)))}dx\eeq
From the study of the pole structure (see section \ref{SectionPole}) we know that $W_{n+1}^{(n-2)}(x.E_{i_0},A)$ has no pole at coinciding points neither at double points, whereas from corollary \ref{Pnkpoles}, the right hand side may have poles only there.
This implies that $W_{n+1}^{(n-2)}(x.E_{i_0},A)dx$ is a meromorphic one-form on the Riemann sphere without any poles. There is no meromorphic differential on the Riemann sphere without poles, except $0$ so that we get:
\beq
W_{n+1}^{(n-2)}(x.E_{i_0},A) =0. \label{Resss}
\eeq
Therefore we conclude that $W_{n+1}^{(n-2)}(x.E_{i_0},A)=0$ i.e. that $W_{n+1}(x.E_{i_0},A)$ is at least of order $\hbar^{n-1}$.

\medskip

We now need to extend the previous result to higher correlators $W_{n+p}$ with $p>1$. For $m\geq n+1$, we define the property $\td{\mathcal{P}}_{n,m}$ :
\begin{center} $\td{\mathcal{P}}_{n,m}$ : $W_m = O(\hbar^{n-1})$ 
\end{center}
We want to prove it by induction on $m\geq n+1$. 

The last result \eqref{Resss} implies that $\td{\mathcal P}_{n,n+1}$ is verified so that initialization of the second induction is done.

Let $m\geq n+1$ and assume that $\td{\mathcal P}_{n,n+1}, \dots, \td{\mathcal P}_{n,m}$ hold. Let $A=\{X_1,\dots,X_m\}$ a set of distinct points of size $m$, and use \eqref{Core}:
\bea \label{Core2} &&P_{m}(x,\y(z^{i_0}(x));A)=\hbar W_{m+1}(x.e_{i_0},A)E_y(x,\y(z^{i_0}(x))) \qquad\cr
&&+\sum_{\{i_0\}\subset I\subset D \,,\,|I|\geq 2}\hbar^{|I|} \hat{\W}_{|I|,m}(I;A)\prod_{i\notin I}(\y(z^{i_0}(x))-\y(z^{i}(x)))
\eea
In the decomposition of definition \ref{defWnhatcal} of $\hat{\W}_{|I|,m}(I;A)$, consider $4$ cases :
\begin{enumerate}\item $|I_i|=1$ and $A_i=\emptyset$ : In that case we get $\hat{W}_{1}(x.E_i)$ which is of order at least $\hbar^0$ (because we have a hat version of $W_1$ whose $\hbar^{-1}$ term is removed).
\item $1<|I_i|+|A_i|\leq n$ : In that case we can apply $\mathcal{P}_{|I_i|+|A_i|}$ and thus we get an order of $\hbar^{|I_i|+|A_i|-2}$
\item $n<|I_i|+|A_i|\leq m$ : In that case, we can apply $\td{\mathcal{P}}_{n,|I_i|+|A_i|}$ and thus we get an order of $\hbar^{n-1}$
\item $|I_i|+|A_i|> m$ : In that case we can only apply $\mathcal{P}_n$ and thus we get an order of $\hbar^{n-2}$
\end{enumerate}
We will denote $L_1$ the set of indexes for which $1<|I_i|+|A_i|\leq n$, $L_2$ the set of indexes for which $n<|I_i|+|A_i|\leq m$ and finally $L_3$ the set of indexes for which $|I_i|+|A_i|> m$. We will also denote $l_1=|L_1|$, $l_2=|L_2|$ and $l_3=|L_3|$. These non-negative integers satisfy $l_1+l_2+l_3=l$. 
Putting it all together, we obtain that $\hat{\W}_{|I|,m}(I;A)$ is of order :
\beq \label{EnCours2}\hbar^{|I|}\prod_{i=1}^l \hat{W}_{|I_i|+|A_i|}(I_i,A_i)=O\left(\hbar^{\underset{i\in L_1}{\sum}(|I_i|+|A_i|-2+\delta_{|I_i|+|A_i|=1})+\underset{i\in L_2}{\sum}(n-1)+\underset{i\in L_3}{\sum}(n-2)+|I| }\right)\eeq
Therefore we need to prove the following inequality :
\beq \label{Inequality2} \underset{i\in L_1}{\sum}(|I_i|+|A_i|-2+\delta_{|I_i|+|A_i|=1})+l_2(n-1)+l_3(n-2)+|I|-n\geq 0\eeq
with $\underset{i=1}{\overset{l}{\sum}}|A_i|=m$ ,$|I_i|\geq 1$, $\underset{i=1}{\overset{l}{\sum}} |I_i|=|I|$ and $l_1+l_2+l_3=l$. 
The case $l_2=l_3=0$ is trivial because we find in \eqref{Inequality2} at least $|I|+m-2l+|I|-n=2(|I|-l)+m-n\geq 0$. In the case $l_2+l_3\geq 0$ we can use the following identity :
\bea 
|I|
&=&\sum_{i\in L_1} |I_i|+\sum_{i\in L_2} |I_i|+\sum_{i\in L_3} |I_i|\cr
&=&\sum_{i\in L_1} |I_i|+\sum_{i\in L_2} (|I_i|+|A_i|)+\sum_{i\in L_3} (|I_i|+|A_i|) - \sum_{i\in L_2\cup L_3} |A_i| \cr
&\geq& l_1 +(n+1)l_2+l_3(m+1) -m
\eea
Inserting this inequality back into \eqref{Inequality2} we find :
\bea &&\underset{i\in L_1}{\sum}(|I_i|+|A_i|-2+\delta_{|I_i|+|A_i|=1})+l_2(n-1)+l_3(n-2)+|I|-n\cr
&&\geq l_2(n-1)+l_3(n-2)+l_1 +(n+1)l_2+l_3(m+1)-m-n\cr
&& \geq 2n l_2 + l_3(n+m-1)-n-m+l_1 \cr
&& \geq 2n l_2 + (l_3-1)(n+m-1)+l_1-1 \cr
&& \geq 2n (l_2+l_3-1) +l_1-1 
%
\eea
If $l_2+l_3>1$ or $l_1>0$, this is clearly non-negative. The only problematic case could be when $l_2+l_3=1$ and $l_1=0$. In this case, there is only one part. This implies that $|A_i|=m$, and $|A_i|+|I_i|=m+|I|>m$ and thus we are in the case $l_3=1$ and $l_2=0$. In this case, inequality \eqref{Inequality2} amounts to $n-2+|I|-n = |I|-2\geq 0$. It is obviously true because the terms with $|I|\leq 1$ are the first line of \eqref{Core2} and have been put aside. Consequently, inequality \eqref{Inequality2} is proved. 

Inserting \eqref{Inequality2} into \eqref{EnCours2}, we deduce that $\hbar^{|I|} \hat{\W}_{|I|,j_0}(I;A)$ is at least of order $\hbar^n$. Since $\hat {\mathcal P}_{n,m}$ holds, we know that $W_{m+1}^{(n-2)}(x.e_{i_0},A)$ is at most of order $O(\hbar^{n-2})$. Writing \eqref{Core2} at order $\hbar^{n-1}$ gives :
\beq \label{Contradiction2}
P_{m}^{(n-1)}(x,\y(z^{i_0}(x));A)\,\frac{dx}{E_y(x,y(z^{i_0}(x)))}=W_{m+1}^{(n-2)}(x.e_{i_0},A) dx
\eeq
Then, the same argument used for \eqref{Contradiction} (i.e. the r.h.s. and the l.h.s. are meromorphic one-forms on the Riemann sphere without any common poles, so they identically vanish) concludes that $W_{m+1}^{(n-2)}(x.e_{i_0},A)=0$. Note that this last statement is only valid when the genus of the curve is zero.

\medskip

Thus we have proved that if $\td{\mathcal{P}}_{n,j}$ is valid for all $n+1\leq j\leq m$ then $\td{\mathcal{P}}_{n,m+1}$ is verified. Since we have proved that $\td{\mathcal{P}}_{n,n+1}$ is also verified we conclude by induction on $m$ that for all $m\geq n+1$, $\td{\mathcal{P}}_{n,m}$ is valid. In other words, for all $m\geq n+1$ : $W_{m+1}(x.e_{i_0},A)=0$ is at least of order $\hbar^{n-1}$. This is precisely proposition $\mathcal{P}_{n+1}$.

\medskip

We finally conclude by induction on $n$ that proposition $\mathcal{P}_n$ is valid for all $n\geq 1$, i.e. that the correlation functions $W_n$ are at least of order $\hbar^{n-2}$.
$\square$

\section{Conclusion \label{S5}}
\subsection{Summary of the results}

We have generalized the proof of \cite{P2,IwakiMarchal} to higher rank systems. We showed that all Lax pairs obeying some assumptions satisfy the TT property, and thus their correlators $W_n$ have an $\hbar$-expansion given by the topological recursion. This result typically lies in a mirror symmetry statement : showing that the A-model correlation functions coincide with the B-model.

We expect that the assumptions we made to prove the TT property, are in fact satisfied by most integrable systems. Our strongest assumption is probably the genus zero assumption, but among integrable systems that have a genus zero spectral curve, we do not know for the moment any example that does not satisfy our assumptions.

If the genus of the spectral curve happens to be strictly positive, then the notions of TT property, WKB expansion and of topological recursion would fail all together.
However, allowing oscillatory terms like in \cite{EynMath} should cure the problem and should give a generalization of the present article. The precise statement of the conjecture is made in \cite{BE10} and the conjecture is strongly supported by the fact that it correctly gives the Jones polynomials to the first few orders in $\hbar$, as verified in \cite{BEknots}.

\subsection{Conjecture for the reconstruction of $\Psi$ via the topological recursion}

So far, we have proved that the determinantal correlation functions $(W_n)_{n\geq 1}$ built from a solution $\Psi$ of the differential system, satisfy the topological recursion.

The next question of interest in quantum curve theory is to ask for the following : how can we recover $\Psi$ from the topological recursion correlation functions $W_{g,n}$'s?
The formulas conjectured in \cite{EOFg, BE10} are the following :

\begin{conjecture}[Exponential formula]
We should have the following WKB expansion : 
\beq
\left(\frac{\Psi(x')^{-1} \Psi(x)}{x-x'}\,\sqrt{dx dx'}\right)_{j,i}\,
= \frac{e^{\frac{1}{\hbar}\int_{z^j(x')}^{z^i(x)} \omega_{0,1}}}{E(z^i(x),z^j(x'))}\,\,e^{\underset{{2g-2+n>0}}{\sum} \frac{\hbar^{2g-2+n}}{n!}\int_{z^j(x')}^{z^i(x)}\dots \int_{z^j(x')}^{z^i(x)} \omega_{g,n}}
\eeq
where $E(z,z')=\frac{(z-z')}{\sqrt{dzdz'}}$ is the Riemann sphere's prime form.

Consequently, the WKB expansion of $\Psi$ should be :
\beq
\Psi(x)_{k,1;i}\,
= \frac{e^{\frac{1}{\hbar} \Phi_k(z^i(x))}}{z^i(x)-\alpha_k}\,\,e^{\underset{{2g-2+n>0}}{\sum} \frac{\hbar^{2g-2+n}}{n!}\int_{\alpha_k}^{z^i(x)}\dots \int_{\alpha_k}^{z^i(x)} \omega_{g,n}}
\eeq
and if $1\leq j\leq d_k$ :
\beq
\left(v^{-1}\Psi(x)\right)_{k,j;i}\,
=  \frac{d^{j-1}}{dz'^{j-1}} \left(\frac{e^{\frac{1}{\hbar} \Phi_k(z^i(x))}}{z^i(x)-z'}\,\,e^{\sum_{2g-2+n>0} \frac{\hbar^{2g-2+n}}{n!}\int_{z'}^{z^i(x)}\dots \int_{z'}^{z^i(x)} \omega_{g,n}} \right)_{z'=\alpha_k}
\eeq
where $\Phi_k(z)$ is a regularized version of $\int_{\alpha_k}^z \omega_{0,1}$ (which is divergent), defined by
\bea
V_k(z) &=& \Res_{z'\to \alpha_k} \omega_{0,1}(z')\,\ln\left(1-\frac{\x(z)^{1/d_k}}{\x(z')^{1/d_k}}\right)
\qquad , \qquad
t_k = \Res_{\alpha_k} \omega_{0,1}\cr
\Phi_k(z) &=& \int_{\alpha_k}^z \left(\omega_{0,1}-dV_k + \frac{t_k}{d_k}\,\frac{d\x}{\x}\right)\,+V_k(z) - \frac{t_k}{d_k} \ln\x(z)
\eea
In other words we define $V_k$ and $t_k$ as the polar part of $\omega_{0,1}$, so that $\omega_{0,1}-dV_k+\frac{t_k d\x}{d_k \x}$ has no pole at $\alpha_k$, we integrate it from $\alpha_k$ to $z$, and add back the term we have subtracted.
\end{conjecture}

\br Note that for any generic point $q$ in a neighborhood of $\alpha_k$, $\Phi_k(z)$ is a regularization of $\int^z \omega_{0,1}$, by adding a constant :
\beq
\Phi_k(z) = C_{q,\alpha_k}+\int_q^z\omega_{0,1} 
\eeq
where $C_{q,\alpha_k}$ is a constant independent of $z$, it depends only on $q$ and $\alpha_k$.
\er

\br
Those formula are of course to be understood in the sense of formal $\hbar$-series.
\er

\br
As usual, those formulas make sense only within open domains where the $z^i(x)$'s, the square roots $\sqrt{\x'(z)}$ and the branches of $\ln\x(z)$ and $\x(z)^{1/d_k}$ are defined. These domains can be called ``Stokes sectors''. They are not global, as is usual with WKB asymptotics implying the Stokes phenomenon : they change when changing sector.
\er

\br
Those formulas should hold only for spectral curves of genus zero, as argued in \cite{BE10,EOFg}. The higher genus formulas are given in \cite{BE10, EynMath}, and \cite{BEknots} is a check for the Jones polynomials of some knots.
\er

It is easy to see that the first few orders in $\hbar$ of those formulas are the right ones. In \cite{BE09, BE10}, it was verified that the conjecture is true in general to order $O\left(\hbar^2\right)$ (i.e. the third non trivial order, since the leading order is $\hbar^{-1}$). 

The main question is to prove that the whole series is indeed formally correct to all order in $\hbar$. The conjecture has been proved to hold for a number of examples : the Airy case proved in \cite{BE09}, the Catalan case in \cite{Penkava}, and many other cases in \cite{MS12,MS15,DM14,Quantum}. Recently, a larger class of examples or rank greater than two was proved in \cite{BouchardEynard2016}.
What is missing at the moment is a general proof that could tackle all orders in a sufficiently generic way.

\section*{Acknowledgments}

B.E. thanks Centre de Recherches Math\'ematiques de Montr\'eal, the FQRNT grant from the Qu\'ebec government. O.M. would like to thank Universit\'e de Lyon, Universit\'e Jean Monnet and Institut Camille Jordan for material support. O.M. work was partially supported by the LABEX MILYON (ANR-10-LABX-0070) of Universit\'e de Lyon, within the program ``Investissements d'Avenir'' (ANR-11-IDEX-0007) operated by the French National Research Agency (ANR). This work is also supported by the ERC Starting Grant no. 335739 ``Quantum fields and knot homologies'' funded by the European Research Council under the European Union's Seventh Framework Programme. B.E. also thanks Piotr Su\l kowski. The authors would also like to thank the organizers of the ``Moduli spaces, integrable systems, and topological recursions'' workshop in Montr\'eal where the first part of this work was realized. They also thank the AMS and the organizers of the ``Von Neumann symposium'' in North Carolina 2016 where part of this work was also realized.

\appendix
\numberwithin{equation}{section}

\section{Recovering $\Psi$ from $M$\label{appequWKBPsiM}}

By definition, $M(x.e_a)=\Psi(x)e_a\Psi(x)^{-1}$ satisfies the ODE :
\beq \label{r}
\hbar\partial_x M(x) = [L(x),M(x)],
\eeq
Moreover, $M(x.e_a)$ is a rank one projector. Let us denote $\Psi(x)^{-1} = \Phi(x)^T$. We have :
\beq
M_{i,j}(x.e_a) = \Psi_{i,a}(x) \Phi_{j,a}(x)
\eeq
and thus, $\forall\, 1\leq k\leq d$ :
\beq \label{rr}
\Psi_{i,a}(x) = {M_{i,k}(x.e_a)}{f_{k,a}(x)} \text{  where  } f_{k,a}(x) = \frac{1}{\Phi_{k,a}(x)}.
\eeq
Let us insert this into the ODE for $\Psi_{i,a}(x)$ :
\beq \label{rrr}
\hbar \partial_x \Psi_{i,a}(x) = \sum_{j=1}^d L_{i,j} \Psi_{j,a}(x)
\eeq
This gives inserting \eqref{rr} into \eqref{rrr} and using \eqref{r} for the derivative of $M$ :
\beq
 [L(x),M(x.e_a)]_{i,k} f_{k,a}(x)  + M_{i,k}(x.e_a) \, \hbar \partial_x f_{k,a}(x)  = (L(x)M(x.e_a))_{i,k} f_{k,a}(x)
\eeq
and thus
\beq
\hbar \frac{\partial_x f_{k,a}(x)}{f_{k,a}(x)}= \frac{(M(x.e_a)L(x))_{i,k} }{ M_{i,k}(x.e_a) },
\eeq
where we notice that the l.h.s. is independent of $i$. Therefore, $\forall\,1\leq i'\leq d$ :
\beq
f_{k,a}(x) = e^{\frac{1}{\hbar}\,\int^x \underset{j=1}{\overset{d}{\sum}} \frac{M_{i',j}(x.e_a) L_{j,k}(x)}{M_{i',k}(x.e_a)}}
\eeq
and $\forall\, 1\leq i,a,k,i'\leq d$ :
\beq
\Psi_{i,a}(x) = M_{i,k}(x.e_a) \,\, e^{\frac{1}{\hbar}\,\int^x \underset{j=1}{\overset{d}{\sum}} \frac{M_{i',j}(x.e_a) L_{j,k}(x)}{M_{i',k}(x.e_a)}}.
\eeq
For example we could chose $i'=k=1$ :
\beq
\Psi_{i,a}(x) = M_{i,1}(x.e_a) \,\, e^{\frac{1}{\hbar}\,\int^x \underset{j=1}{\overset{d}{\sum}} \frac{M_{1,j}(x.e_a) L_{j,1}(x)}{M_{1,1}(x.e_a)}}.
\eeq
This can also be written
\beq
\Psi_{i,a}(x) = \frac{M_{i,k}(x.e_a)}{M_{i',k}(x.e_a)} \,\, e^{\frac{1}{\hbar}\,\int^x \underset{j=1}{\overset{d}{\sum}} L_{i',j}(x)\,\frac{ M_{j,k}(x.e_a) }{M_{i',k}(x.e_a)}}.
\eeq
In particular, choosing $i'=i$ :
\beq
\Psi_{i,a}(x) =  \,\, e^{\frac{1}{\hbar}\,\int^x \underset{j=1}{\overset{d}{\sum}} L_{i,j}(x)\,\frac{ M_{j,k}(x.e_a) }{M_{i,k}(x.e_a)}}.
\eeq
Similarly
\beq
\Phi_{i,a}(x) =  \,\, e^{\frac{-1}{\hbar}\,\int^x \underset{j=1}{\overset{d}{\sum}} L_{j,i}(x)\,\frac{ M_{k,j}(x.e_a) }{M_{k,i}(x.e_a)}}.
\eeq

In conclusion, if both $M(x.e_a)$ and $L(x)$ have a formal $\hbar$ power series expansion, then $\Psi$ has a WKB expansion. Note that the converse is also true from theorem \ref{Mexp} : if $L(x)$ has a power series expansion in $\hbar$ and $\Psi(x)$ has a WKB expansion then $M(x.e_a)$ has a power series expansion in $\hbar$.

\section{Examples : Painlev\'{e} and $(p,q)$ minimal models \label{AppendixExamples}}

In this appendix, we present various cases in which our method can be applied. The first one deals with $(p,q)$ minimal models that were studied in \cite{BBE14}. The second one deals with the Painlev\'{e} Lax pairs and was developed in \cite{IwakiMarchal}. For clarity we will only focus on the Painlev\'{e} VI case though all other Painlev\'{e} systems can be treated similarly (details can be found in \cite{IwakiMarchal}). The purpose of this section is also to give interesting examples for which all assumptions presented in this paper are satisfied.

\subsection{$(p,q)$ minimal models}

These were studied with the topological recursion in \cite{BBE14}. However the proof presented in \cite{BBE14} was incomplete (the proof of the leading order property used an insertion operator. A part of the definition of this operator was missing. The gap was completed for $q=2$ in \cite{BookEynard} but the general case remained incomplete). This new proof doesn't use insertion operators, it uses our general loop equations method. We will here follow the standard notations of \cite{BBE14} taking in particular $q=d$.

In $(p,q)$ minimal models ($p$ and $q$ are coprime strictly positive integers, see \cite{BBE14} for details), $R(x,t,\hbar)$ is a $q\times q$ companion matrix :
\beq
R(x,t,\hbar) = \begin{pmatrix}
0 & 1 & 0 & \dots & 0 \cr
 & 0 & 1 & & \vdots \cr
 \vdots & & & \ddots & 0\cr
0 & & \dots & 0 & 1\cr
 u_{d-1}(t,\hbar) & & \dots & u_1(t,\hbar) & u_{0}(t,\hbar)-x
\end{pmatrix}
\eeq
The matrix $\Psi(x,t)$, described in \cite{BBE14}, is given by :
\beq \Psi(x,t,\hbar)=\begin{pmatrix} \psi_1(x,t)& \dots&\psi_q(x,t)\\
(\hbar\partial_t)\psi_1(x,t)&\dots&(\hbar \partial_t)\psi_q(x,t)\\
\vdots&&\vdots\\
(\hbar \partial_t)^{q-1}\psi_1(x,t)&\dots&(\hbar\partial_t)^{q-1}\psi_q(x,t)
\end{pmatrix}
\eeq
where $(\psi_i)_{1\leq i\leq q}$ are linearly independent solutions of the system :
\beq x\psi(x,t)=Q\psi(x,t) \text{  ,  } \hbar \partial_t \psi(x,t)=-P\psi(x,t) \text{  and  } [P,Q]=\hbar\eeq
where the operator $(P,Q)$ are of the form :
\bea P&=&\sum_{k=0}^p v_k(t)(\hbar\partial_t)^k, \,\,\,\, v_p=1, v_{p-1}=0, v_{p-2}=-p u(t)\cr
Q&=&\sum_{l=0}^q u_l(t)(\hbar\partial_t)^l, \,\,\,\, u_q=1, u_{q-1}=0, u_{p-2}=-q u(t)
\eea
In particular, the condition $[P,Q]=\hbar$ determines all functions $(v_i)_{1\leq i\leq p}$ and $(u_i)_{1\leq i\leq q}$ in terms of $u(t)$ and its derivatives. The $L(x,t)=\left(L_{k,j}(x,t)\right)_{1\leq k,j\leq q}$ matrix is determined by decomposing the operators $(L_k)_{k\geq 0}$ on the basis $\left((\hbar \partial_t)^i\right)_{i\geq 0}$ :
\beq L_k(x,t)=\sum_{j=0}^q L_{k,j}(x,t) (\hbar \partial_t)^j\eeq
where the operators $(L_k)_{k\geq 0}$ are defined recursively as :
\beq L_0(x,t)=-\sum_{l=0}^p v_l(t) F_l(x,t) \,\, ,\,\, L_{k+1}(x,t)=(\hbar \partial_t)L_k(x,t)+L_{k,q-1}(x,t)(x-Q)\eeq 
with $F_l(x,t)=\underset{j\geq 0}{\sum} F_{l,j}(x,t)(\hbar \partial_t)^j$ defined recursively by :
\beq F_0(x,t)=1 \,\,,\,\, F_{l+1}(x,t)= (\hbar \partial_t)F_l(x,t)+F_{l,q-1}(x,t)(x-Q)\eeq
In particular, it is obvious from the definitions that $L(x,t)$ is a polynomial in $x$.

\medskip
In the context of $(p,q)$ minimal models, one is interested in formal expansion in $\hbar$. Since the functions $\left(u_i(t,\hbar)\right)_{i\geq 0}$ and $\left(v_i(t,\hbar)\right)_{i\geq 0}$ admit a formal expansion in $\hbar$, we get that assumption \ref{Asshbarexpansion} is verified. Moreover, the spectral curve is of genus $0$, so assumption \ref{genuszeroassumption} is verified. It is given by (see Proposition $5.2$ of \cite{BBE14}) :
\bea \label{SpecCurvepq} \x_t(z)&=&\sum_{k=0}^q u_k^{(0)}(t)z^k\cr
\y_t(z)&=&\sum_{l=0}^p v_l^{(0)}(t)z^l
\eea
The auxiliary spectral curve is given by the characteristic polynomial of the companion matrix $R(x,t,\hbar)$ :
\beq
\tilde E(x,s,t,\hbar=0) = \det(s - R^{(0)}(x,t)) = \x_t(s)-x
\eeq
The set of solutions of  $\tilde E_t(x,s;t,0)=0$ is thus the set of all $(\x_t(z),z)$ for $z\in \curve=\bar{\mathbb C}$. Therefore the auxiliary spectral curve is equivalent to the triple : 
\beq
\tilde{\mathcal S}_t = (\bar{\mathbb C},\x_t,\s_t)
\eeq
with the function $\s_t$ is the identity map $\s_t : z\mapsto z$. Obviously the auxiliary spectral curve does not admit any double points and the spectral curve \eqref{SpecCurvepq} is regular so assumption \ref{Assnodblpt} is verified. Note that in our setting, the poles of the $\x_t$ function correspond to $k=1$, $d_1=q$ and $\alpha_1=\infty$. In other words, $z\mapsto \x_t(z)$ has only one pole at infinity of order $q$ (in the general theory developed above, the point $z=\infty$ was assumed not to be a pole of $\x$. This means that some of the above formulas require some basic adaptations to accommodate this particular case). Since the $R(x,t)$ matrix is a companion matrix, its eigenvectors are given by a Vandermonde-like matrix and we obtain :
\beq V(x,t)=\mathcal{V}(x) \,\,\,\Rightarrow \,\,\, v(t)=I_q\eeq
In particular, assumption \ref{AssumptionV} is trivially satisfied.

\medskip

Notice that by definition, $L(x,t)$ is a polynomial in $x$ whose coefficients admit an $\hbar$-expansion. Thus, assumption \ref{Asspolesordrek} is satisfied.

\medskip

Assumption \ref{Asssym} was partly proved in \cite{BBE14}. Indeed, the authors proved that the matrix $\Gamma(t)$ given by (note that there is a change of convention in \cite{BBE14} where the $\Gamma$ matrix is defined as the inverse of our present matrix and with a global $(-1)^{q-1}$ constant) :
\beq \Gamma(t)=\gamma(t)^{-1} \text{  with  } \gamma(t)=(-1)^{q-1}\Phi(x,t)\Psi(x,t)^T\eeq
satisfy \eqref{Parity1} (See \cite{BBE14} for a precise definition of $\Phi(x,t)$). In particular Theorem $5.2$ of \cite{BBE14} proves that the matrix $\gamma(t)$ does not depend on $x$. Therefore the only remaining issue to prove assumption \ref{Asssym} is to match $\Gamma^{(0)}$ with $\left(v(t)^T\right)^{-1}C v(t)^{-1}$ to satisfy \eqref{Gamma0}. We observe that by definition, the generalized Vandermonde matrix $\mathcal{V}(x)$ leads to :
\beq C=\begin{pmatrix}u_1^{(0)}(t)& u_2^{(0)}(t)&\dots &u_{q-2}^{(0)}(t)&0&1\\
u_2^{(0)}(t)&\iddots &\iddots &\iddots&1&0\\
\vdots& \iddots&\iddots&\iddots&\iddots&0\\
u_{q-2}^{(0)}(t)&\iddots&\iddots&\iddots& \iddots&\vdots\\
0&1&\iddots&\iddots&\iddots&0\\
1&0&0& \dots &0&0
\end{pmatrix}  
\eeq
In other words : $C_{i,j}=0$ if $i+j>q+1$ and $C_{i,j}=u^{(0)}(t)_{i+j-1}$ if $i+j\leq q+1$. Its inverse is given by :
\beq C^{-1}=\begin{pmatrix}0& 0&\dots &0&0&1\\
0&\iddots &\iddots &\iddots&1&a_2\\
\vdots& \iddots&\iddots&\iddots&\iddots&a_3\\
0&\iddots&\iddots&\iddots& \iddots&\vdots\\
0&1&\iddots&\iddots&\iddots&a_{q-1}\\
1&a_2&a_3&\dots& a_{q-1}&a_q
\end{pmatrix} 
\eeq
In other words, $\left(C^{-1}\right)_{i,j}=0$ if $i+j< q+1$ and $\left(C^{-1}\right)_{i,j}=a_{i+j-q}$ if $i+j\geq q+1$. The coefficients $(a_i)_{1\leq i\leq q}$ are determined by the following recursion (obtained by looking at the term $(C^{-1}C)_{i,1}=\delta_{i,1}$ with $1\leq i\leq q$)  :
\beq \label{CA} a_1=1\,\,,\,\, a_2=0\,\,\, \text{ and }\,\, a_{i+1}=-\sum_{j=1}^{i-1} a_j u_{j+q-i+1}^{(0)}(t) \text{ for } 2\leq i\leq q-1\eeq
Since $v(t)=I_q$, condition \eqref{Gamma0} is equivalent to prove that $C^{-1}=\gamma^{(0)}(t)$. The matrix $\gamma(t)$ (unfortunately denoted $\textbf{C}$  with entries labeled from $0$ to $q-1$ in \cite{BBE14}) is described in equations $5.77$, $5.78$ and $5.79$ of \cite{BBE14}. It satisfies $\gamma_{i,j}=0$ if $i+j<q+1$ and
\bea  \gamma_{1,j}&=&\delta_{j,q} \text{  for  } 1\leq j\leq q\cr
\hbar \partial_t \gamma_{i,j}&=& \gamma_{i,j+1}-\gamma_{i+1,j} \text{  for  } 1\leq i,j\leq q-1\cr
\hbar \partial_t \gamma_{i,q-1}&=&-\gamma_{i,q}-\sum_{l=0}^{q-2} u_l(t)\gamma_{i,l+1} \text{  for  } 1\leq i\leq q-1
\eea
Let us denote for clarity $B=\gamma^{(0)}$. Projecting the last set of equations at order $\hbar^0$ gives $B_{i,j}=0$ if $i+j<q+1$ and :
\bea \label{SetB} B_{1,j}&=& \delta_{j,q} \text{  for  } 1\leq j\leq q\cr
B_{i+1,j}&=& B_{i,j+1} \text{  for  } 1\leq i,j\leq q-1\cr
B_{i,q}&=&-\sum_{l=0}^{q-2} u_l^{(0)}(t)B_{i,l+1} \text{  for  } 1\leq i\leq q-1
\eea
The second equation is equivalent to say that $B$ is a Hankel matrix of the same form as $C^{-1}$. In other words, $B_{i,j}=0$ if $i+j<q+1$ and $B_{i,j}=b_{i+j-q}$ if $i+j\geq q+1$. The coefficients $(b_i)_{1\leq i\leq q}$ are determined by the first and last equations of \eqref{SetB}. We get :
\beq b_1=1 \,\,,\,\, b_2=0\,\,\text{ and }\,\, b_{i+1}=-\sum_{l=1}^{q-2} b_l u_{l}^{(0)}(t) \text{ for } 2\leq i\leq q-1\eeq 
Hence we recover the same recursion as \eqref{CA}. This finally proves that $C^{-1}=\gamma^{(0)}$ so that assumption \ref{Asssym} is verified.

\medskip

In conclusion, we have proved all required assumptions for the $(p,q)$ minimal models that therefore satisfy the Topological Type property.

\subsection{Painlev\'e VI case}

Painlev\'{e} equations were studied with the topological recursion in \cite{P2} (Painlev\'{e} II) and \cite{IwakiMarchal} (all six Painlev\'{e} equations). A simpler method (only valid in the case $d=2$) was used to prove that the Painlev\'{e} Lax pairs satisfy the topological type property. We propose here to show that our generalization also applies directly to these cases. We will only carry out the Painlev\'{e} VI case (which is the most difficult) in details but all results presented here can be easily adapted to the other Painlev\'{e} cases using computations presented in \cite{IwakiMarchal}.

\medskip

In the Painlev\'e 6 system we have  :
\beq \label{P6}L_{\rm VI}(x,t,\hbar) = \frac{A_0(t)}{x}+\frac{A_1(t)}{x-1}+\frac{A_t(t)}{x-t}\,\,,\,\, R_{\rm VI}(x,t,\hbar) =-\frac{A_t(t)}{x-t}-\frac{(q-t)(\theta_\infty-\hbar)}{2t(t-1)}\sigma_3 \eeq
\beaa  A_0&=&\begin{pmatrix} z_0+\frac{\theta_0}{2}&-\frac{q}{t}\\ \frac{tz_0(z_0+\theta_0)}{q}&-\left(z_0+\frac{\theta_0}{2}\right)\end{pmatrix} \,\,,\,\, A_1=\begin{pmatrix} z_1+\frac{\theta_1}{2}&\frac{q-1}{t-1}\\ -\frac{(t-1)z_1(z_1+\theta_1)}{q-1}&-\left(z_1+\frac{\theta_1}{2}\right)\end{pmatrix}\cr
A_t&=&\begin{pmatrix} z_t+\frac{\theta_t}{2}&-\frac{q-t}{t(t-1)}\\ \frac{t(t-1)z_t(z_t+\theta_t)}{q-t}&-\left(z_t+\frac{\theta_t}{2}\right)\end{pmatrix}\,\,,\,\,
A_\infty=\begin{pmatrix} \frac{\theta_\infty}{2}&0\\0&-\frac{\theta_\infty}{2}\end{pmatrix}=-(A_0+A_1+A_t)
\eeaa
Here, $z_0(t), z_1(t)$ and $z_t(t)$ are auxiliary functions of $t$ that can be expressed in terms $q(t)$ and a function $p(t)$ defined by :
\beq p=\frac{z_0+\theta_0}{q}+\frac{z_1+\theta_1}{q-1}+\frac{z_t+\theta_t}{q-t}\eeq
The explicit expression for $z_0$, $z_1$ and $z_t$ in terms of $q$ can be found in \cite{IwakiMarchal} where $q(t)$ is shown to satisfy a $\hbar$-deformed version of the Painlev\'{e} $6$ equation (see \cite{IwakiMarchal} for details). Note that the matrix form $L(x,t)dx$ has simple poles at $x\in\{0,1,\infty,t\}$ while $R(x,t)dx$ only has simple poles at $x\in\{\infty,t\}$. Existence of an $\hbar$-expansion is discussed in \cite{IwakiMarchal} where assumption \ref{Asshbarexpansion} is proved. At first order in $\hbar$ it is shown in \cite{IwakiMarchal} that the spectral and auxiliary curves are of genus $0$ :
\bea \label{SpecCurveP6} y^2 &=& \frac{\theta_{\infty}^2(x-q_0)^2 P_2(x)}{4x^2(x-1)^2(x-t)^2}\cr
s^2&=&\frac{(q_0-t)^2\theta_\infty^2P_2(x)}{4t^2(t-1)^2(x-t)^2}
\eea
where $P_2(x)=x^2+\left(-1-\frac{\theta_0^2t^2}{\theta_\infty^2 q_0^2}+\frac{\theta_1^2(t-1)^2}{\theta_\infty^2(q_0-1)^2}\right)x+\frac{\theta_0^2t^2}{\theta_\infty^2q_0^2}=(x-a)(x-b)$ that can be written equivalently $P_2(x)=x^2+\left(-\frac{\theta_0^2t(t+1)}{\theta_\infty^2q_0^2}+\frac{\theta_1^2t(t-1)}{\theta_\infty^2(q_0-1)^2}-\frac{\theta_t^2t(t-1)}{\theta_\infty^2(q_0-t)^2}\right)x+\frac{\theta_0^2t^2}{\theta_\infty^2q_0^2}$. Here $q_0$ stands for $q^{(0)}(t)$ the leading order in $\hbar$ of $q(t)$. It satisfies an algebraic equation of degree $6$ that can be found explicitly in \cite{IwakiMarchal}. Inserting this result in the definition of $R^{(0)}(x,t)$, we get an expression of $z_t^{(0)}$ and $q_0$ in terms of $a, b$ and $t$ (and the monodromy parameters):
\bea \label{AuxP6} 
z_t^{(0)}&=&-\frac{\theta_t}{2}+\frac{1}{4}+\frac{\theta_\infty(q_0-t)\left(t-\frac{a+b}{2}\right)}{2t(t-1)}=-\frac{\theta_t}{2}\pm\frac{t-\frac{a+b}{2}}{2\sqrt{(t-a)(t-b)}}\cr
(q_0-t)&=&\pm\frac{t(t-1)\theta_t}{\theta_\infty\sqrt{(t-a)(t-b)}}
\eea
so that we get :
\bea \label{R0P6} R^{(0)}(x,t)&=&\pm\begin{pmatrix}-\frac{\theta_t\left(x-\frac{a+b}{2}\right)}{2(x-t)\sqrt{(t-a)(t-b)}}& \frac{\theta_t}{\theta_\infty(x-t)\sqrt{(t-a)(t-b)}}\\ -\frac{(b-a)^2\theta_t\theta_\infty}{16(x-t)\sqrt{(t-a)(t-b)}}&  \frac{\theta_t\left(x-\frac{a+b}{2}\right)}{2(x-t)\sqrt{(t-a)(t-b)}}
\end{pmatrix}\cr
L^{(0)}(x,t)&=&\frac{(x-q_0)t(t-1)}{x(x-1)(q_0-t)}R^{(0)}(x,t) 
\eea 
The spectral curve \eqref{SpecCurveP6} is of genus $0$ with two finite branchpoints located at the two simple zeros of the polynomial $P_2$ denoted $a$ and $b$. Thus assumption \ref{genuszeroassumption} is satisfied. Note that there is also a double point at $x=q_0$ for the spectral curve but it is absent in the auxiliary curve. Since the spectral curve is of genus $0$, it can be parametrized globally on $\overline{\mathbb{C}}$ and we choose a parametrization suitable with the convention that $z=\infty$ is not a pole of $x(z)$ (so that it slightly differs from the usual Zhukovski parametrization of \cite{IwakiMarchal}). We take:
\bea \x(z)&=&\frac{a+b}{2}+\frac{b-a}{2}\left(1+\frac{1}{z-1}-\frac{1}{z+1}\right)=b+\frac{b-a}{(z+1)(z-1)}\cr
\y(z)&=&\frac{\theta_\infty(x(z)-q_0)(b-a)z}{2(z-1)(z+1)x(z)(x(z)-1)(x(z)-t)}\cr
\s(z)&=&\frac{(q_0-t)\theta_\infty(b-a)z}{2(z-1)(z+1)t(t-1)(x(z)-t)}\cr
&=&\pm \frac{(b-a)z\theta_t}{2(z-1)(z+1)(x(z)-t)\sqrt{(t-a)(t-b)}}
\eea
Note that $\x'(z)=-\frac{2z(b-a)}{(z+1)^2(z-1)^2}$. In the $z$ variable, the two branchpoints are located at $z=0$ and $z=\infty$ while the poles are located at $z=\pm 1$. The involution (corresponding to $\x(z)=\x(\bar{z})$) is given by $\bar{z}=-z$. Inverting the relation between $x$ and $z$ leads to :
\beq z^1(x)=\sqrt{\frac{x-a}{x-b}} \text{  and  } z^2(x)=-\sqrt{\frac{x-a}{x-b}}\eeq
so that
\bea S_1(x)&=&\frac{\theta_\infty(q_0-t)\sqrt{(x-a)(x-b)}}{t(t-1)(x-t)}=\pm\frac{\theta_t}{(x-t)}\sqrt{\frac{(x-a)(x-b)}{(t-a)(t-b)}}\cr
S_2(x)&=&-\frac{\theta_\infty(q_0-t)\sqrt{(x-a)(x-b)}}{t(t-1)(x-t)}=\mp\frac{\theta_t}{(x-t)}\sqrt{\frac{(x-a)(x-b)}{(t-a)(t-b)}}\cr
Y_1(x)&=&\frac{\theta_\infty(x-q_0)\sqrt{(x-a)(x-b)}}{x(x-1)(x-t)}\cr
Y_2(x)&=&-\frac{\theta_\infty(x-q_0)\sqrt{(x-a)(x-b)}}{x(x-1)(x-t)}
\eea
In particular, from the last identities it is straightforward to verify that the auxiliary curve has no double points, i.e. that assumption \ref{Assnodblpt} is satisfied. Moreover, application of the previous formulas leads to :
\beq \vec{\mathcal V}(z)=\left( -\frac{i(z+1)}{\sqrt{2(b-a)}\sqrt{z}},-\frac{i(z-1)}{\sqrt{2(b-a)}\sqrt{z}}\right)\eeq
and thus :
\bea \label{PainleveV} \mathcal{V}(x)&=&\begin{pmatrix} -\frac{i(z^1(x)+1)}{\sqrt{2(b-a)}}\left(\frac{x-b}{x-a}\right)^{\frac{1}{4}}& -\frac{(z^2(x)+1)}{\sqrt{2(b-a)}}\left(\frac{x-b}{x-a}\right)^{\frac{1}{4}}\\
-\frac{i(z^1(x)-1)}{\sqrt{2(b-a)}}\left(\frac{x-b}{x-a}\right)^{\frac{1}{4}}& -\frac{(z^2(x)-1)}{\sqrt{2(b-a)}}\left(\frac{x-b}{x-a}\right)^{\frac{1}{4}} 
\end{pmatrix}\cr
&=&\begin{pmatrix}
 -\frac{i}{\sqrt{2(b-a)}}\frac{\sqrt{x-a}+\sqrt{x-b}}{\left((x-a)(x-b)\right)^{\frac{1}{4}}}&-\frac{1}{\sqrt{2(b-a)}}\frac{\sqrt{x-b}-\sqrt{x-a}}{\left((x-a)(x-b)\right)^{\frac{1}{4}}}\\
-\frac{i}{\sqrt{2(b-a)}}\frac{\sqrt{x-a}-\sqrt{x-b}}{\left((x-a)(x-b)\right)^{\frac{1}{4}}} &\frac{1}{\sqrt{2(b-a)}}\frac{\sqrt{x-b}+\sqrt{x-a}}{\left((x-a)(x-b)\right)^{\frac{1}{4}}}\end{pmatrix}\cr
&=&\frac{1}{\sqrt{2(b-a)}}\begin{pmatrix} -i\left(\left(\frac{x-a}{x-b}\right)^{\frac{1}{4}}+\left(\frac{x-b}{x-a}\right)^{\frac{1}{4}}\right)&\left(\frac{x-a}{x-b}\right)^{\frac{1}{4}}-\left(\frac{x-b}{x-a}\right)^{\frac{1}{4}}\\
i\left(\left(\frac{x-b}{x-a}\right)^{\frac{1}{4}}-\left(\frac{x-a}{x-b}\right)^{\frac{1}{4}}\right) &\left(\frac{x-a}{x-b}\right)^{\frac{1}{4}}+\left(\frac{x-b}{x-a}\right)^{\frac{1}{4}}\end{pmatrix} 
\eea
It is then straightforward to verify that :
\bea \mathcal{V}(x)\mathcal{V}(x)^T&=&\begin{pmatrix} \frac{(z^2(x)-z^1(x))}{(b-a)}\left(\frac{x-b}{x-a}\right)^{\frac{1}{2}}&\frac{((z^2(x))^2-(z^1(x))^2)}{(b-a)}\left(\frac{x-b}{x-a}\right)^{\frac{1}{2}}\\
 \frac{((z^2(x))^2-(z^1(x))^2)}{(b-a)}\left(\frac{x-b}{x-a}\right)^{\frac{1}{2}}& -\frac{(z^2(x)-z^1(x))}{(b-a)}\left(\frac{x-b}{x-a}\right)^{\frac{1}{2}}
\end{pmatrix}\cr
&=&\begin{pmatrix}-\frac{2}{b-a}&0\\0&\frac{2}{b-a}\end{pmatrix}
\eea
Hence we get $C=\frac{b-a}{2}\text{ diag}(-1,1)$ as claimed from \eqref{EqC}. Note that we also get :
\bea \mathcal{V}(x,t)\begin{pmatrix}1&0\\0&0\end{pmatrix}\mathcal{V}(x,t)^T&=&\begin{pmatrix}-\frac{ \sqrt{\frac{x-a}{x-b}}+\sqrt{\frac{x-b}{x-a}}+2}{2(b-a)}& \frac{ \sqrt{\frac{x-b}{x-a}}-\sqrt{\frac{x-a}{x-b}}}{2(b-a)}\\ \frac{ \sqrt{\frac{x-b}{x-a}}-\sqrt{\frac{x-a}{x-b}}}{2(b-a)}& -\frac{ \sqrt{\frac{x-a}{x-b}}+\sqrt{\frac{x-b}{x-a}}-2}{2(b-a)}\end{pmatrix}\cr
\mathcal{V}(x,t)\begin{pmatrix}0&0\\0&1\end{pmatrix}\mathcal{V}(x,t)^T&=&\begin{pmatrix}\frac{ \sqrt{\frac{x-a}{x-b}}+\sqrt{\frac{x-b}{x-a}}-2}{2(b-a)}& -\frac{ \sqrt{\frac{x-b}{x-a}}-\sqrt{\frac{x-a}{x-b}}}{2(b-a)}\\ -\frac{ \sqrt{\frac{x-b}{x-a}}-\sqrt{\frac{x-a}{x-b}}}{2(b-a)}& \frac{ \sqrt{\frac{x-a}{x-b}}+\sqrt{\frac{x-b}{x-a}}+2}{2(b-a)}\end{pmatrix}
\eea
Computing $\mathcal{V}(x)S(x)\mathcal{V}(x)^TC$ leads to :
\small{\bea &&\mathcal{V}(x)S(x)\mathcal{V}(x)^TC=\begin{pmatrix}\frac{\theta_\infty(q_0-t)}{2t(t-1)}+ \frac{\theta_\infty(q_0-t)\left(t-\frac{a+b}{2}\right)}{2(x-t)}&-\frac{\theta_\infty(q_0-t)(b-a)}{4t(t-1)(x-t)}\\
\frac{\theta_\infty(q_0-t)(b-a)}{4t(t-1)(x-t)}& -\frac{\theta_\infty(q_0-t)}{2t(t-1)}- \frac{\theta_\infty(q_0-t)\left(t-\frac{a+b}{2}\right)}{2(x-t)}
 \end{pmatrix}\cr
&&=\pm\begin{pmatrix}\frac{\theta_t}{\sqrt{(t-a)(t-b)}}+\frac{\left(t-\frac{a+b}{2}\right)t(t-1)\theta_t}{2\sqrt{(t-a)(t-b)}(x-t)}& -\frac{\theta_t(b-a)}{4\sqrt{(t-a)(t-b)}(x-t)}\\ \frac{\theta_t(b-a)}{4\sqrt{(t-a)(t-b)}(x-t)}&-\frac{\theta_t}{\sqrt{(t-a)(t-b)}}-\frac{\left(t-\frac{a+b}{2}\right)t(t-1)\theta_t}{2\sqrt{(t-a)(t-b)}(x-t)}\end{pmatrix}
\eea}
\normalsize{}where we used \eqref{AuxP6} to replace $q_0$. Eventually a direct computation from \eqref{PainleveV} and \eqref{R0P6} shows that :
\bea V(x,t)&=&v(t)\mathcal{V}(x) \text{ with } v(t)=\begin{pmatrix} 0&\frac{4}{\theta_\infty(b-a)} \\ 1&0\end{pmatrix}\cr
L^{(0)}(x,t)&=&v(t)\mathcal{V}(x)Y(x)\mathcal{V}(x)^TCv(t)^T \cr
R^{(0)}(x,t)&=&v(t)\mathcal{V}(x)S(x)\mathcal{V}(x)^TCv(t)^T 
\eea
so that assumption \ref{AssumptionV} is verified.

\medskip

Eventually since 
\bea L_{\text{VI}}(x,t,\hbar) &=& \frac{A_0(t,\hbar)}{x}+\frac{A_1(t,\hbar)}{x-1}+\frac{A_t(t,\hbar)}{x-t}\,\,,\cr
 R_{\text{VI}}(x,t,\hbar) &=&-\frac{A_t(t,\hbar)}{x-t}-\frac{(q-t)(\theta_\infty-\hbar)}{2t(t-1)}\sigma_3 
\eea
we see that there is no mixing between the $x$-dependence and the $\hbar$-expansion. In particular, $L^{(k)}$ has poles only at $x\in\{0,1,t\}$ and assumption \ref{Asspolesordrek} is trivially verified. Finally, the symmetry condition is answered in \cite{IwakiMarchal} where it is proved that 
\beqq \Gamma_{\text{VI}}(t)=\begin{pmatrix}-\frac{t^2 z_0(z_0+\theta_0)}{q}+\frac{(t-1)^2z_1(z_1+\theta_1)}{q-1}&0\\0&1 \end{pmatrix}\eeqq
satisfies assumption \ref{Asssym}. Note that at order $\hbar^0$ computations from \cite{IwakiMarchal} gives :
\beq \Gamma^{(0)}_{\text{VI}}(t)=\begin{pmatrix}-\frac{\theta_\infty^2(b-a)^2}{16}&0\\0&1 \end{pmatrix}\eeq
Since $\Gamma$ is only determined up to a global multiplication by a constant, we can easily match it with the direct computation of :
\beq (v^T)^{-1} C v^{-1}=\frac{b-a}{2}\begin{pmatrix}-\frac{\theta_\infty^2(b-a)^2}{16} &0\\ 0&1\end{pmatrix}\eeq
and thus assumption \ref{Asssym} is satisfied.

\end{document}